# Integrated Microwave Photonics:
# From Material Platforms to Systems-on-Chip

Jiejun Zhang[1,2] and Jianping Yao[3*]

*1. College of Physics & Optoelectronic Engineering, Jinan University, Guangzhou 511443, China.*

*2. International Cooperation Joint Laboratory for Optoelectronic Hybrid Integrated Circuits, Jinan University, Guangzhou 511443, China*

*3. Microwave Photonics Research Laboratory, School of Electrical Engineering and Computer Science, University of Ottawa, Ottawa, ON K1N 6N5, Canada.*

E-mail: jpyao@uottawa.ca

## Abstract

Microwave photonics (MWP) is a rapidly evolving interdisciplinary field that bridges photonics and microwave engineering, harnessing the unique strengths of photonic devices and circuits to generate, transport, process, and measure high-frequency signals in the optical domain for applications across the microwave, millimeter-wave, and sub-terahertz spectral regimes. Integrated microwave photonics (IMWP) represents the next evolutionary step within this field, where the MWP functionalities are not realized using bulky, costly, and alignment-sensitive discrete photonic components, but instead are miniaturized and densely integrated onto compact, chip-scale photonic platforms and modules. This integration enables smaller, more stable, less power-consuming, and more scalable MWP modules for applications, such as radar, wireless communication, spectrum sensing, and analog signal processing. In this paper, recent advances in IMWP are reviewed, including material platforms, integration technologies, and system functionalities. Emerging opportunities and future perspectives are discussed. Silicon (Si) and silicon nitride (SiN) provide dense routing, programmable filtering, and low-loss delay; indium phosphide (InP) supplies optical gain, light generation, and high-speed photodetection; thin-film lithium niobate (TFLN) offers linear and broadband electro-optic conversion; and photonic integration is emerging as a practical path to synergize complementary capabilities. Recent work shows a marked shift from isolated modulators, filters, and delay lines toward chip- and module-level systems, ranging from signal processing engines, real-time spectrum sensing, to full-spectrum wireless links, fiber-wireless conversion, silicon beamforming, integrated radar, programmable processors, and photonic convolution engines with on-chip sources and detectors. As these demonstrations become more mature and complete, the key questions move beyond bandwidth and efficiency toward system integration, microwave packaging, calibration stability, analog link performance, and application-level validation.

## 1. Introduction

Originating from early studies of microwave-optical interactions, microwave photonics (MWP) leverages photonic principles to address bandwidth bottlenecks, signal degradation, and electromagnetic susceptibility in conventional microwave systems, thereby enhancing their performance and capabilities for modern applications. By mapping microwave information onto an optical carrier or carriers, microwave functions can be performed in the optical domain, which enable the exploitation of the wide bandwidth, low transmission loss, immunity to electromagnetic interference (EMI), and wavelength multiplexing capability, thereby realizing microwave functions that are difficult to implement in the electrical domain [1], [2], [3]. The established functional categories include photonic-assisted microwave signal generation, MWP signal processing, true-time-delay beamforming, radio-over-fiber transmission, optoelectronic oscillation, instantaneous frequency measurement, and MWP radar and sensing. These validated functionalities demonstrate the significant potential of MWP for practical applications; however, the most powerful implementations of these functions currently rely on discrete components, which greatly limits the widespread adoption of MWP technology [1], [2], [3].

Integrated microwave photonics (IMWP) was motivated by the need to overcome the inherent limitations of discrete MWP systems, particularly in terms of size, weight, power consumption, environmental sensitivity, and assembly complexity [4], [5]. By enabling integration of key photonic and electronic components onto a single platform, IMWP significantly reduces system footprint and improves stability, while also enhancing scalability and reproducibility for large-scale deployment. An earlier review article by Marpaung et al. framed IMWP as a route toward chip-scale MWP processors with multifunctionality and reconfigurability [4]. Since then, two closely linked developments have reshaped the field. First, advances in material platforms have enabled a complementary set of building blocks: silicon (Si)  and silicon nitride (SiN) for dense passive circuits and low-loss routing, indium phosphide (InP) for optical gain and light generation, and high-speed photodetection, thin-film lithium niobate (TFLN) for linear and ultrabroadband electro-optic conversion, and heterogeneous or hybrid integration for functional partitioning [6]. Second, rapid progress in system-level demonstrations has shifted the focus from individual components toward fully functional MWP systems, moving the evaluation criteria beyond delay lines, filters, and modulators toward complete MWP functionalities [7], [8], [9], [10], [11], [12], [13], [14].

The discussion below adopts a system-oriented perspective. SOI, SiN, InP, TFLN, ferroelectric materials, and heterogeneous integration, each occupies a different region of the design space. Packaging and co-design are treated as central topics, as they are essential for the realization of complex system-on-chip architectures. The survey focuses on recent advances in IMWP for microwave frontend, radar, spectrum sensing, wireless communications, and computing engines [8], [9], [10], [12], [15], [16].

A chronological map of the most influential and highly cited milestones used to structure this review is provided in Fig. 1. This paper is organized as follows. In Section 2, photonic integration platforms and analog microwave-link metrics are discussed and compared. In Section 3, heterogeneous or hybrid integration, MWP packaging, and electronic co-design are discussed. In Section 4, a survey on integrated MWP system demonstrations is provided. In Section 5, the roles of different platforms are assessed from a system-level perspective. Finally, in Section 6, deployment challenges and reporting metrics for future work are summarized.

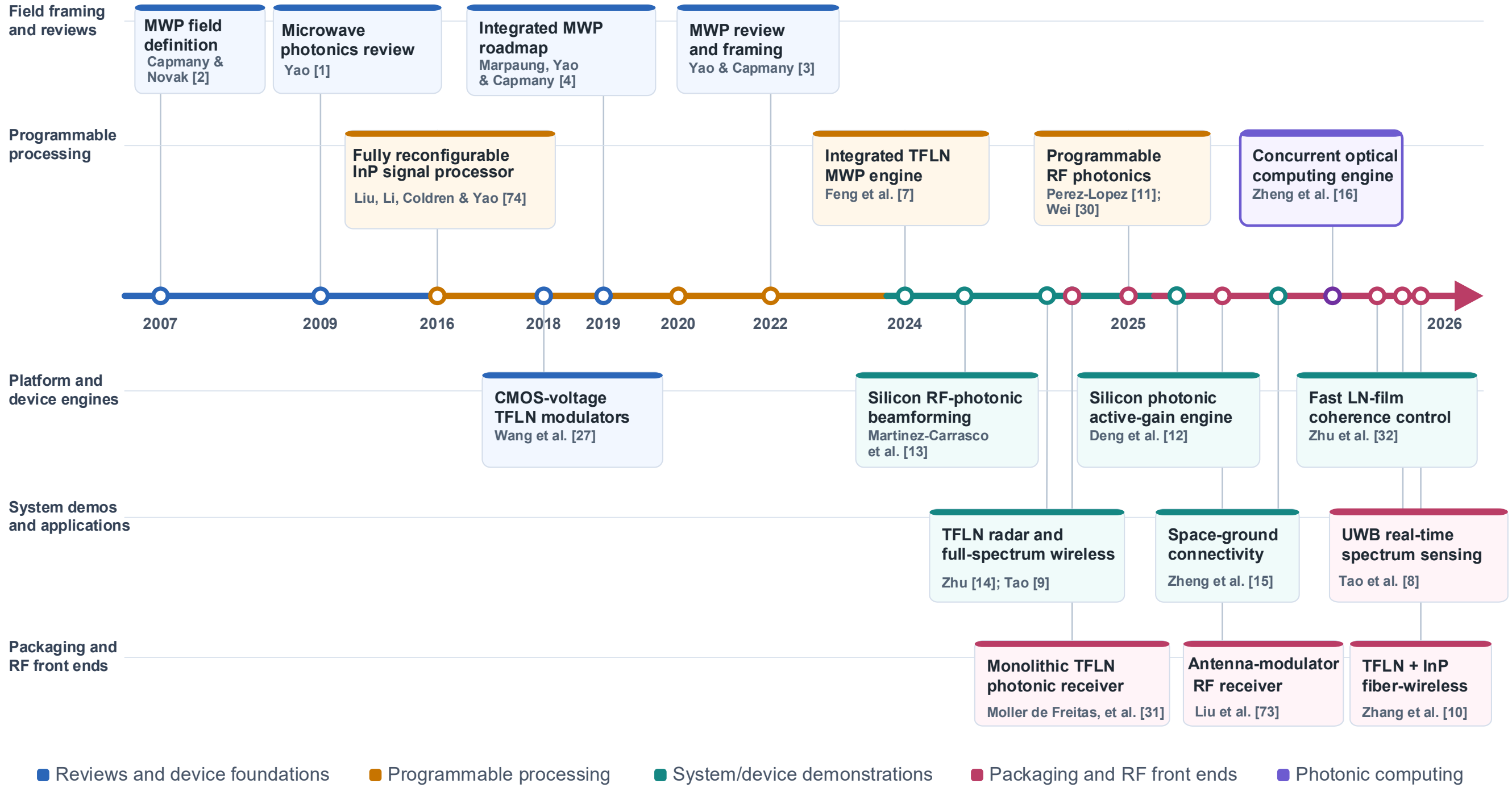

*Fig. 1. Key milestones in IMWP. The timeline summarizes representative works that have shaped the field, organized by reviews and device foundations, programmable processing, platform and device engines, system-level demonstrations, packaging and RF front ends, and photonic computing. Recent years have witnessed an accelerating convergence across several key domains, including programmable photonic processors, TFLN and silicon photonic platforms, wireless/radar demonstrations, and monolithic integration of RF-photonic receivers.*

# 2. Material Platforms for IMWP

## 2.1 Requirements for Integrated Platforms

Platform choice in IMWP is governed by a small set of microwave-relevant functions: optical carrier generation, microwave-to-optical conversion, optical time delay and filtering, wavelength routing, optical-to-microwave detection, and microwave/electronic packaging. Each function stresses distinct requirements on material properties. Low-loss delay lines require minimal scattering and absorption; high-dynamic-range links require highly linear modulation and high-saturation-power detection; reconfigurable processors require stable tuning and monitoring; system-level demonstrations depend critically on robust packaging and accurate calibration. Table 1 provides a summary of the main platform roles from this functional perspective.

*Table 1. Comparison of major IMWP material platforms.*

| Platform | Main strengths | Main limitations | Typical IMWP roles | Representative references |
|---|---|---|---|---|
| SOI | Foundry scale; dense routing; Ge PDs | No native gain; carrier nonlinearity | Switching, filtering, delay, beamforming | [11], [12], [13], [17], [18], [19] |
| SiN | Very low loss; high power handling | No native gain; weak EO/OE activity | Delay, filtering, combs, interposers | [20], [21], [22] |
| InP | Lasers; SOAs; high-speed PDs | High passive loss; low density | Sources, gain, mmWave/THz detection | [23], [24], [25], [26] |
| TFLN | Pockels modulation; high linearity; broad bandwidth | No native gain; high thermo-optic sensitivity; packaging immaturity | EO front ends, radar, sensing, processing | System-level integration [7], [8], [9], [10], [30], [31]; platform reviews and modulators [27], [28], [29] |
| BTO/ferroelectrics | Large Pockels effect; Si integration prospects | Process immaturity; low domain stability | Future low-voltage EO conversion | [38] |

## 2.2 Silicon and Silicon Nitride Platforms

Silicon photonics, realized primarily on the silicon on insulator (SOI) platform, has been the dominant platform for large-scale photonic integrated circuits (PICs) because it benefits from mature CMOS fabrication, compact high-index-contrast waveguides, dense wavelength routing, germanium-on-silicon photodetection, and an increasingly developed foundry ecosystem [17], [18]. For IMWP, the major advantage of silicon is its large scalability. Complex switched optical delay lines, microring filters, multiport meshes, wavelength-selective processors, and beamforming networks can be fabricated with high component counts and repeatable layout rules. Recent silicon signal processors have combined low-loss waveguides, low-phase-error optical switches, and wavelength or time  interleavers to support beamforming, filtering, arbitrary waveform generation, and parallel optical computing on a single scalable chip [19]. MWP beamforming based on silicon has also advanced to ultrabroadband and high-resolution designs, with integrated delay line networks addressing the beam squint problem existing in conventional phase-shifter based array antennas [13].

The intrinsic limitations of SOI impose significant constraints on the design and implementation of IMWP systems. Unlike direct-bandgap materials such as gallium arsenide (GaAs) or indium phosphide (InP), silicon is an indirect-bandgap material and cannot produce optical gain or support lasing in its native form. To overcome this inherent deficiency, on-chip light sources must rely on heterogeneous or hybrid integration with a direct-bandgap material. This integration, whether achieved through epitaxial growth, wafer bonding, or flip-chip assembly, introduces additional fabrication complexity, thermal management challenges, and cost overhead, yet it remains an essential enabler for fully functional silicon photonic circuits that require on-chip laser sources and optical amplifiers. In addition, silicon high-speed modulators rely on the plasma-dispersion effect, in which carrier-induced changes in the refractive index are inherently accompanied by changes in optical absorption, resulting in limited modulation linearity and increased optical loss. Although these limitations are generally acceptable for digital optical communications, analog MWP links are more sensitive to link gain, noise figure, third-order intermodulation distortions (IMDs), and spurious-free dynamic range (SFDR). Consequently, these fundamental constraints have driven the development of several innovative solutions aimed at circumventing the plasma-dispersion trade-offs inherent to silicon modulators. One class of approaches focuses on modulator linearization, employing predistortion, feedforward, or feedback techniques to mitigate the nonlinear distortion introduced by carrier-induced effects. A second avenue involves hybrid silicon/III–V modulators, which integrate III–V compound semiconductors, with their superior electro-absorption or electro-optic coefficients, onto silicon waveguides, thereby achieving higher modulation efficiency and improved linearity. A third, and increasingly compelling, direction is the heterogeneous integration of Pockels-effect materials, particularly thin-film lithium niobate (TFLN) and barium titanate (BTO). These materials offer a linear electro-optic response that is intrinsically decoupled from optical absorption, promising high-linearity, low-loss modulation that is well suited for demanding analog applications.

SiN has emerged as a complementary platform to SOI, offering a distinct set of passive optical properties that include ultralow optical loss and high optical power handling capability. These attributes are particularly advantageous for applications such as long time delays, optical filtering, and microcomb generation [20], [21], [22]. However, its centrosymmetric structure precludes a native Pockels effect, while its weak Kerr nonlinearity is insufficient for practical electro-optic modulation. Consequently, SiN cannot independently perform high-speed optical modulation or generate optical gain and lasing. To overcome these limitations, SiN is most effectively utilized in heterogeneous or hybrid integration schemes, where it is combined with InP-based gain media for light generation, TFLN for high-speed electro-optic modulation, and silicon for photodetection. In practical IMWP systems, SiN commonly serves as a low-loss passive routing layer or optical interposer, providing ultralow-loss signal distribution and spectral processing while delegating active functions to complementary material platforms.

## 2.3 Indium Phosphide Active Photonic Integration

InP is widely regarded as the most complete monolithic platform for active photonic integration. Unlike silicon, which suffers from an indirect bandgap that precludes efficient light emission, InP is a direct-bandgap semiconductor that enables high-efficiency optical gain and light generation. InP supports a rich suite of active optoelectronic functionalities on a single epitaxial structure, including lasers, semiconductor optical amplifiers (SOAs), electro-absorption modulators (EAMs), Mach-Zehnder modulators (MZMs), and photodetectors (PDs), all of which can be monolithically integrated without the need for heterogeneous bonding or hybrid assembly [23]. This capability is strategically important for IMWP, as a practical and deployable MWP system cannot rely indefinitely on multiple off-chip lasers, amplifiers, and other discrete components. InP also remains essential for high-speed and high-power photodetection. For example, charge-compensated modified uni-traveling-carrier photodiodes (UTC-PDs) have enabled mmWave generation at frequencies up to 325 GHz with strong output power performance [24]. More recently, waveguide-integrated UTC-PDs have achieved bandwidths of approximately 140 GHz with very low dark current [25]. These capabilities are particularly relevant to emerging applications, including D-band and sub-THz communications, photonic-assisted microwave local oscillators, and high-power and wideband MWP links.

The primary limitation of InP is its relatively poor passive scalability. InP waveguides exhibit lower index contrast and higher propagation losses compared to SOI, SiN, or TFLN, making dense passive routing, long delay lines, and complex filter networks more difficult to implement. Additionally, InP fabrication relies on costly epitaxial growth and regrowth steps, which are less amenable to the high-volume, planar processing used in silicon photonics. As a result, while InP offers unrivalled active functionality, it is less suited for large-scale passive optical networks, where SOI, SiN and TFLN maintain clear advantages in routing density, loss, and cost-effectiveness. Although InP based SOAs can be used to compensate for loss, amplified spontaneous emission (ASE) and nonlinear gain dynamics introduce additional noise and distortions, which are undesirable in low-noise and high-dynamic-range microwave links. As a result, InP is often better viewed as an active photonic layer or chiplet-based light sources, such as lasers, SOAs, and high-speed photodiodes, rather than the host for implementing extensive passive and programmable photonic circuitry. For example, an InP-LiNbO3 wafer-level coherent receiver achieving 3.584 Tbps demonstrates the effectiveness of integrating InP photodiodes and lithium niobate modulators to realize high-speed communication links [26].

### 2.4 Thin-Film Lithium Niobate Electro-Optic Integration

TFLN has recently emerged as a central platform for IMWP, largely because it offers a uniquely compelling solution to one of the most persistent challenges in modern MWP systems: the microwave-to-optical conversion bottleneck. Its Pockels effect enables fast, linear, and low-loss phase modulation without free-carrier absorption. Integrated lithium niobate modulators operating at CMOS-compatible driving voltages were demonstrated in 2018 [27], and subsequent studies have established TFLN as an effective platform for highly linear and wide-bandwidth electro-optic photonics [28], [29]. For MWP, these properties are particularly important, since they can translate into high analog linearity, large electro-optic bandwidth, low drive voltage, and strong compatibility with mmWave and sub-THz systems, thereby supporting efficient, high-performance signal generation, modulation, transmission, and processing in advanced IMWP systems.

Recent system-level demonstrations have further shown that TFLN can be extended beyond single-functional and standalone modulator components to realize photonic integrated subsystems. A wafer-scale TFLN MWP processing engine was demonstrated which could perform temporal integration and differentiation at a speed up to 256 GSa/s, with a processing bandwidth up to 67 GHz, and was applied for differential equation solving, ultrawideband waveform generation, and image edge detection [7]. A programmable multifunctional TFLN MWP circuit integrating an intensity modulator with multiple cascaded microring resonators has been applied to achieve tunable notch filtering and effective broadband interference suppression [30]. A monolithically integrated antenna-coupled TFLN photonic receiver has been demonstrated, in which both microwave antenna structures and high-speed photonic modulators are co-fabricated on a single TFLN-on-quartz platform. This co-integration eliminates the need for discrete antenna-to-modulator interconnects, reducing parasitic losses and form factor while preserving the broad electro-optic bandwidth of TFLN, thereby enabling compact, high-performance ultrawideband wireless front-end operation [31]. Recently, a TFLN real-time spectrum sensing chip has been demonstrated that integrates three key functional blocks on a single platform, a broadband MZM, an electro-optic microring filter bank, and an electro-optic comb reference. This monolithic integration enables real-time detection and characterization of signals spanning from microwave to sub-THz frequencies, while maintaining a processing latency of less than 110 ns [8]. A more recent demonstration of high-speed optical coherence manipulation using a TFLN modulator shows that a single electro-optic platform can provide rapid phase- and coherence-control functions essential for programmable delay, filtering, and coherent microwave-photonic processing [32].

TFLN also underpins recent full-spectrum and ultrawideband wireless demonstrations. In 2025, a TFLN photonic wireless system was reported that operates across an exceptionally broad frequency range from 0.5 to 115 GHz. The system supports adaptive wireless communications across nine independent bands, with each band capable of sustaining data rates up to 100 Gbps, showcasing the potential of TFLN for future high-capacity, multi-band wireless networks [9]. In 2026, a fully photonic integrated system for ultrawideband fiber-wireless communication was demonstrated, combining TFLN electro-optic conversion and InP modified UTC photodiodes. The TFLN modulator provides broad electro-optic bandwidth and low drive voltage, while the InP photodiodes offer efficient high-power photodetection across mmWave and sub-THz frequencies. Jointly, they achieve a system bandwidth exceeding 250 GHz, supporting 512 Gbps fiber transmission and 400 Gbps wireless links [10]. These results position TFLN among the pioneering platforms that have pushed IMWP beyond the modulator level and into true application-level demonstrations.

Despite its many advantages, the principal remaining weakness of TFLN is the lack of optical gain. Through rare-earth doping, however, TFLN can provide optical gain. For example, rare-earth-doped TFLN waveguide amplifiers, including Er-doped and Er-Yb-co-doped, have been reported to achieve on-chip gain [33], [34], but an optical pumping source is needed. Heterogeneous III-V-on-lithium-niobate integration, which bonds gain material directly onto the TFLN waveguide layer, and photonic-wire-bonded InP/TFLN external-cavity lasers, which employ polymer waveguides to connect discrete gain chips to passive TFLN circuits, both offer practical, electrical pumping source solutions that circumvent the limitations of rare-earth-doped amplifiers [35], [36], [37]. However, low-noise, thermally stable, yield-compatible laser/amplifier integration for deployable IMWP modules remains considerably less mature than the well-established TFLN modulation platform. Near-term TFLN systems are expected to adopt hybrid approaches that leverage discrete InP lasers and detectors, or co-packaged external optical sources, until fully integrated gain solutions reach comparable technological readiness.

## 2.5 Emerging Ferroelectric Electro-Optic Platforms

While TFLN has emerged as the leading platform for Pockels electro-optic photonics, a broader family of ferroelectric oxide materials is now attracting increasing research interest. Barium titanate (BTO), thin-film lithium tantalate (TFLT), and related ferroelectric oxides are being actively explored as complementary or alternative platforms, each offering distinct advantages that may extend the capabilities of integrated electro-optic devices. The central question is whether their intrinsic and thin-film material properties can support reproducible, broadband, and low-voltage electro-optic conversion. BTO, for instance, is especially attractive from this perspective because strained and microstructured BTO films exhibit an exceptionally high effective Pockels coefficient that could enable compact phase shifters and low-voltage modulators [38]. Recent monolithic BTO modulators on SOI substrates further demonstrate that effective electro-optic modulation can be obtained from epitaxial BTO films, although the measured effective coefficients, optical loss, and device variability remain strongly linked to film orientation, crystalline quality, strain state, and domain structure [39]. In addition, the demonstrations of cryogenic BTO modulators also indicate that the material response can remain useful in low-temperature environments, which is relevant to quantum-photonic and superconducting-electronic interfaces [40]. All these results position BTO as a high-potential platform for low-voltage electro-optic conversion; however, its relevance at the system level is contingent upon the successful resolution of several material-level challenges, namely the wafer-scale film uniformity, domain stability under bias and temperature cycling, optical loss after etching, RF dielectric loss, and long-term reproducibility.

Other ferroelectric films such as PZT provide a valuable point of comparison for the broader field of electro-optic photonics. They demonstrate that strong Pockels activity can be introduced into other passive photonic platforms, such as SiN, opening new possibilities for active functionality on established photonic circuits. However, they also highlight the persistent trade-offs that define the electro-optic material landscape, higher electro-optic coefficients often come at the cost of increased optical loss, while process compatibility and long-term reliability remain challenging to achieve simultaneously [41]. In an IMWP context, these materials should therefore be viewed less as immediate replacements for the now-dominant TFLN platform and more as specialized electro-optic materials that may offer distinct advantages in targeted applications. They are most likely to find value where the primary constraints are the drive voltage, footprint, operation at cryogenic temperatures, or the need for heterogeneous integration with a specific passive platform such as SiN or SOI.

TFLT is a different case from BTO or PZT, because $LiTaO_3$ is structurally and chemically much closer to $LiNbO_3$, sharing the same crystal family and analogous Pockels coefficients, while also offering a practical advantage: a mature industrial ecosystem already exists for $LiTaO_3$ in the form of radio-frequency acoustic filters, which provides established supply chains, manufacturing experience, and potential for rapid scaling. Recent lithium tantalate PICs have demonstrated low-loss $LiTaO_3$ waveguides, MZMs, soliton microcomb generators, and compatibility with deep-ultraviolet lithography, making TFLT attractive from a manufacturability perspective [42]. Compared with TFLN, $LiTaO_3$ offers lower birefringence, which can simplify broadband and dense circuit design, and its established wafer supply chain may eventually reduce substrate-cost barriers. High-speed TFLT modulators have also reached ultrabroadband operation for data-communication links, showing that the material can support the needed bandwidths for electro-optic front ends [43]. Earlier work on $LiTaO_3$ microdisk reporting a high optical damage threshold is also relevant for resonant or high-optical-power photonic systems [44]. Nevertheless, TFLT still lags behind TFLN in terms of mature foundry access, low-loss component libraries, microwave-electrode design rules, and system-level MWP demonstrations. Its most realistic near-term role is as a manufacturability-oriented ferroelectric platform that could complement TFLN if low-loss routing, broadband modulation, RF packaging, and wafer-scale process repeatability converge in integrated MWP systems.

### 2.6 Analog-Link Metrics and Platform Partitioning

For digital photonics, the key performance metrics differ significantly from those of analog MWP systems. The primary figures of merit for digital photonics are insertion loss, bandwidth, extinction ratio, energy per bit, and yield. For IMWP, the relevant figure set is considerably wider. The microwave link gain depends on optical carrier power, modulator efficiency, photodetector responsivity, microwave impedance matching, and all passive losses between them. Noise figure depends not only on photodetector shot noise but also on laser relative intensity noise, amplifier noise, optical insertion loss before detection, and the noise introduced by any optical gain element. Linearity depends on the electro-optic transfer function, modulator bias, photodetector saturation, optical filtering, and microwave packaging. These metrics are usually framed using analog optical-link theory [45], [46], while chip-scale filters, delay lines, and waveform processors provide the circuit examples discussed later [47], [48], [49], [50], [51], [52], [53]. A platform that is excellent for digital transceivers can still be unattractive for analog microwave links if the modulator or detector introduces strong nonlinear distortions.

The contrast between SOI and TFLN is especially instructive. SOI can realize dense programmable circuits and robust foundry workflows. However, the reliance on carrier-depletion modulators introduces inherent nonlinearities in both phase and amplitude, requiring compensatory measures such as bias control, segmented electrodes, linearization circuits, or digital predistortion to achieve acceptable analog performance [17], [18], [19], [54], [55], [56]. TFLN offers an intrinsically linear Pockels response and can maintain high electro-optic bandwidth with low optical loss, directly accounts for its rapid adoption in recent demonstrations of radar systems, wireless links, and spectrum sensing [7], [8], [9], [10], [30], [31]. InP remains indispensable for optical gain and high-speed photodetection, while passive delay and filtering networks are often more efficiently implemented in low-loss platforms such as SOI, SiN, or TFLN [23], [24], [25], [26], [57].

The platform comparison is most useful when it is interpreted as functional partitioning, a recognition that no single material platform optimizes every function in a MWP system. A practical MWP module may integrate multiple platforms: a III-V gain coupon on silicon for the laser source; a TFLN or BTO modulator for high-linearity microwave-to-optical conversion; SiN or SOI low-loss passive routing, filtering, and delays; and Ge or InP photodiodes for detection, with the choice determined by the specific frequency and saturation current requirements of the application [12], [38], [57], [58], [59].

## 3. Photonic Integration and Packaging

The platform comparison discussed above makes heterogeneous and hybrid integration hard to avoid, as no single material system can simultaneously satisfy the diverse demands of active gain, passive routing, electro-optic modulation, and scalable manufacturing [60]. Silicon offers the manufacturing scale essential for cost-effective production, InP delivers gain and high-speed detection required for active signal processing, TFLN provides linear electro-optic conversion necessary for low-distortion modulation, and SiN provides ultralow-loss passive routing that preserves signal integrity over complex waveguide networks. The central practical challenge, however, lies not in selecting one platform over another, but in synergistically combining these complementary strengths into a cohesive system without sacrificing microwave performance, thermal stability, optical coupling efficiency, and manufacturability at scale. This imperative co-engineering demands that optical chiplets and waveguide layers be designed together with microwave ports, package transitions, thermal dissipation paths, control electronics, and calibration software, each of which introduces its own constraints and trade-offs that must be reconciled through iterative and cross-domain optimization.

### 3.1 Wafer-Scale Integration

Heterogeneous integration typically refers to integration of dissimilar materials onto a common photonic wafer platform through techniques such as the bonding or transfer printing, which enables functionality that no single material could provide alone. III-V-on-silicon integration represents the most mature example of this approach, allowing InP gain sections to be coupled into underlying silicon waveguides using adiabatic taper structures that gradually transition the optical field distribution from the III-V layer to the silicon layer [57]. This approach has already revolutionized datacom transceiver industry, leading to the massive deployment of silicon-photonic optical engines in data centers worldwide. It has begun to attract significant attention in the IMWP community, because MWP system-on-chip require stable, low-noise, and precisely controlled on-chip optical carriers. For example, a single-chip silicon photonic engine reported by Deng et al. provides a compelling illustration of this direction [12]. The chip combines transfer-printed InP optical gain sections with silicon modulators, tunable filters, photodetectors, optical switches, and microwave interfaces. At the system boundary, the internal optical implementation is abstracted behind microwave and optical interfaces: input microwave signals are converted, processed optically on chip, and converted back to microwave outputs. This integration transitions MWP away from laboratory-scale optical assemblies and toward more practically deployable MWP systems.

### 3.2 Die-Level Hybrid Assembly and Photonic Wire Bonding

Die-level hybrid assembly offers flexibility in cases where wafer-scale process compatibility is difficult to achieve. Flip-chip bonding enables the attachment of lasers, SOAs, TFLN modulators, or photodetectors onto a passive interposer, while photonic wire bonding uses three-dimensional polymer waveguides fabricated by two-photon lithography to connect waveguides across chips with relaxed alignment constraints [58]. These approaches are particularly valuable for IMWP modules where optimal lasers, modulators, delay network, and detectors may come from different foundries.

For IMWP, the packaging problem is not restricted to optical coupling. Microwave launch structures, impedance matching, thermal dissipation paths, ground-signal-ground transitions, bias routing, and control electronics must be co-designed with the photonic circuit, drawing on multi-chip photonic assembly, photonic wire bonding, and silicon-photonic coupling/interposer technologies [58], [61], [62], [63]. At mmWave and sub-THz frequencies, package parasitics can erase the bandwidth advantage of the photonic devices, rendering microwave transitions and board- or package-level electromagnetic design a core component of the IMWP research problem [9], [10], [31].

### 3.3 Microwave-Photonic-Electronic Co-Design

A large-scale IMWP engine demands more than just the photonic chip itself; it requires control electronics, calibration algorithms, and in many cases DSP or machine-learning-based equalization. Recent demonstrations of programmable processors have shown that software layers, monitoring photodetectors, thermal controllers, and calibration routines are indispensable for rendering large photonic circuits practically useful [11], [12], [64]. The next generation of IMWP systems will require co-design spanning electromagnetic simulation, photonic circuit design, electronic packaging, thermal modeling, and control algorithms.

## 4. Integrated MWP Systems

### 4.1 System Evidence and Classification Criteria

In this section, a system is treated as an IMWP system when the principal MWP function is implemented on a photonic integrated platform or within a co-packaged photonic subsystem, and when the system is tested at a meaningful application boundary. Such systems are separated into three practical categories. In a PIC, the photonic circuit executes the central microwave function, such as time delay, filtering, modulation, spectrum mapping, or waveform processing. In a hybrid integrated module, multiple material platforms, microwave interfaces, and photodetectors are assembled to demonstrate an application-level function. In a transferable concept, the reported work validates a MWP function but leaves substantial source integration, packaging, calibration, or microwave interfacing to future implementations.

Table 2 summarizes recent IMWP system demonstrations, highlighting a clear trend toward functional partitioning across platforms, with TFLN emerging as the platform of choice for ultrafast signal processing, radar, and wireless links, SOI well suited for beamforming and compact transceivers, and heterogeneous integration (SOI+InP) delivering full engine-level functionality on a single chip.

*Table 2. Recent representative integrated IMWP system demonstrations.*

| Year | Work | Integration status | Platform | Integrated function | Main system evidence |
| --- | --- | --- | --- | --- | --- |
| 2024 | Ultrabroadband high-resolution silicon microwave photonic beamformer [13] | PIC | SOI | Switched optical true-time-delay beamforming | Eight optical delay lines and five delay stages for broadband phased-array operation |
| 2024 | Integrated TFLN MWP processing engine [7] | Wafer-scale PIC | TFLN | Temporal integration, differentiation, and analog waveform processing | Up to 67 GHz processing bandwidth and up to 256 GSa/s operation |
| 2024 | Dual-band integrated MWP transceiver for SAR [65] | PIC with radar validation | SOI | Radar waveform generation and stretch processing | C/Ku-band reconfigurable operation and SAR imaging validation |
| 2025 | Antenna-coupled ultrawideband photonic receiver [31] | Integrated receiver chip | TFLN-on-quartz | Antenna-to-optical microwave capture | Monolithic antenna-photonic receiver for ultrawideband wireless front-end operation |
| 2025 | Integrated LN photonic mmWave radar [14] | PIC with radar validation | TFLN | mmWave waveform generation and de-chirping | 40-50 GHz V-band radar and 1.50 cm ranging resolution |
| 2025 | Programmable multifunctional TFLN MWP circuit [30] | PIC | TFLN | Tunable notch filtering and interference suppression | Programmable microring-based microwave photonic filtering on TFLN |
| 2025 | Single-chip silicon photonic engine [12] | Heterogeneously integrated engine | SOI + InP | Source, modulation, filtering, detection, switching, and OEO functions | 5 mm x 1.3 mm core with on-chip light source and microwave/optical interfaces |
| 2025 | Full-spectrum wireless communications [9] | Photonic integrated wireless subsystem | TFLN | Carrier/LO generation, modulation, and wireless-photonic conversion | 0.5-115 GHz operation and up to 100 Gbps lane speed |
| 2026 | Real-time spectrum sensing for 6G [8] | Integrated sensing chip | TFLN | MZM, EO comb, and EO microring filter bank | 57.5 GHz analysis bandwidth, up to 120 GHz measurable frequency, and <110 ns latency |
| 2026 | Integrated fiber-wireless communication [10] | Hybrid integrated EO/OE link components | TFLN + InP MUTC-PD | EO/OE conversion for converged fiber-wireless links | 512 Gbps fiber and 400 Gbps wireless transmission |
| 2025 | Space-ground dual-band IMWP transceiver [15] | Monolithically integrated SOI transceiver | SOI | Microwave TTD, optical phased array, and coherent optical transceiver | 10 Gbps microwave and 80 Gbps/wavelength optical free-space links over 5 m |

### 4.2 Beamforming and Microwave-Optical Transceiving on Integrated Platforms

True time delay (TTD) beamforming is one of the most natural applications of IMWP. In addition to eliminating beam squint over a wide microwave bandwidth, photonic integrated implementations offer excellent scalability, enabling large-scale beamforming networks with compact footprints. Furthermore, integrated optical delay lines are inherently stable, providing highly accurate and repeatable time delays for broadband beamforming systems. Traditional electronic phased arrays use phase shifters, where the phase delay translates into frequency-dependent time delays, leading to beam squint over wide bandwidths. In contrast, photonic delay lines implement true physical time delay, thereby maintaining a fixed beam direction over a broad bandwidth.

In 2024, Martinez-Carrasco et al. proposed a resolution-improved switched optical true-time-delay architecture and demonstrated its implementation on an SOI chip, consisting of eight optical true-time-delay lines and five delay stages, as shown in Fig. 2 [13]. This work serves as a representative integrated demonstration of silicon-photonic TTD beamforming architectures, highlighting the feasibility of scalable on-chip delay networks. It addresses the scalability challenge of binary switched delay networks, where higher angular resolution typically demands more switches, increased footprint, and higher insertion loss. The proposed design demonstrated how dense passive integration of silicon photonics can support high-resolution MWP beamforming.

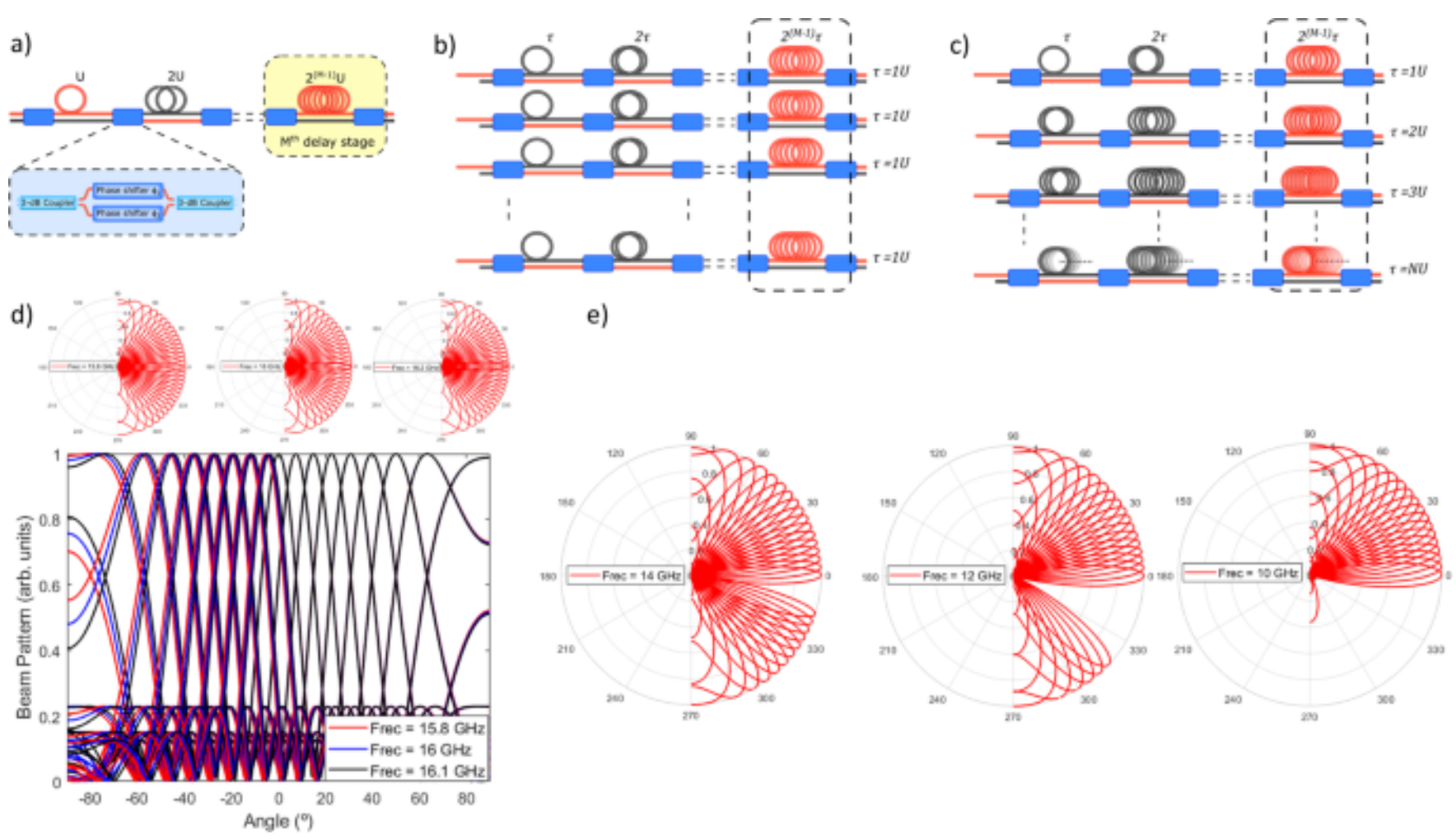

*Fig. 2. Silicon photonic true-time-delay beamformer design. The figure compares switched true-time-delay line designs and the beam-squint limitation that motivates equalized delay-line approaches. Source: Martinez-Carrasco et al., "Ultrabroadband high-resolution silicon RF-photonic beamformer," Nature Communications 15, 1433 (2024), Ref. [13], https://doi.org/10.1038/s41467-024-45743-9.*

In 2025, Zheng et al. extended the beamforming concept from microwave arrays to a dual-band architecture integrating microwave arrays and optical arrays, enabling simultaneous microwave and free-space optical transceiving [15]. As shown in Fig. 3, the dual-band transceiver was implemented on a monolithically integrated SOI chip that combines a microwave true-time-delay beamforming network, an optical phased array network, and an optical coherent transceiver. In a reported 5 m link, 10 Gbps microwave transmission and 80 Gbps per wavelength near-infrared transmission were demonstrated, with dynamic beam steering in both bands. The relevance of the work to this review lies in the co-integration of microwave phased-array beamforming, optical phased-array beam steering, and coherent transceiving functions on a single SOI platform.

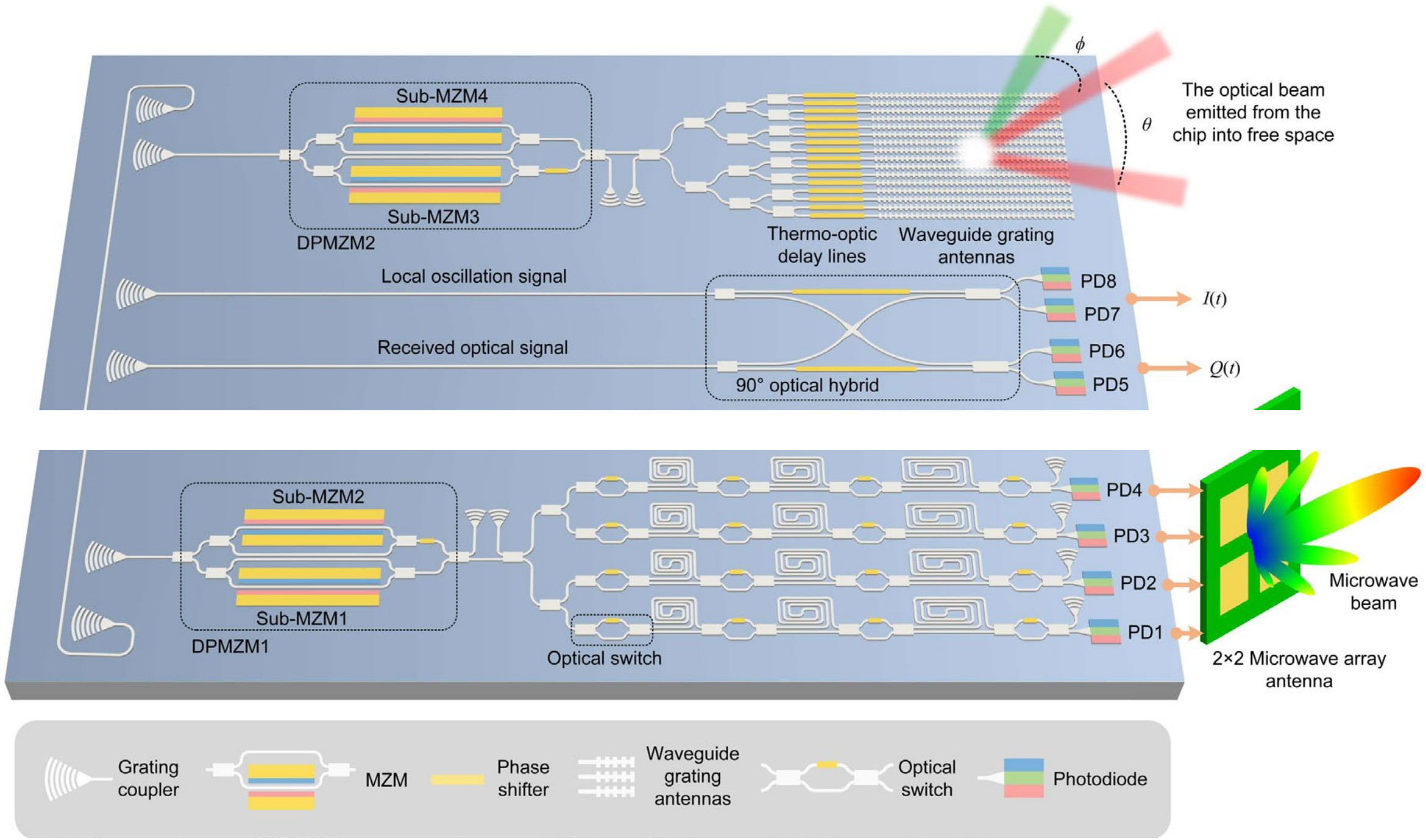

*Fig. 3. Monolithically integrated dual-band SOI transceiver for space-ground connectivity. The schematic shows the key system blocks of the SOI transceiver: dual-parallel MZM for microwave signal generation, a thermo-optic true-time-delay network, a microwave phased array, optical switches, an optical phased array, a coherent optical receiver, and balanced PDs. Source: Zheng et al., "Microwave photonics for space-ground connectivity," arXiv:2509.18018, 2025, Ref. [15], https://doi.org/10.48550/arXiv.2509.18018.*

Beyond demonstrating fundamental functionality, these beamforming studies also experience critical integration challenges. While optical true-time-delay networks can mitigate beam squint and scale to a large number of channels, their competitiveness against electronic phased arrays hinges on practical RF link metrics. Key determinants such as link gain, insertion loss, photodetector saturation, antenna co-integration, and thermal calibration must be rigorously addressed. Future integrated beamforming studies should address not only fundamental optical parameters such as delay resolution and insertion loss, but also the system-level performance that ultimately determine practical viability. These include microwave-frequency link gain and noise figure, which govern the overall signal fidelity, as well as photodetector linearity and saturation behavior under realistic optical power levels. Equally critical are array-level calibration strategies, which must account for element-to-element phase and amplitude variations across the full operational bandwidth and over extended temperature ranges. Furthermore, packaging assumptions, often relegated to footnotes, deserve explicit treatment, as they profoundly influence thermal management, interconnect parasitics, antenna coupling, and the feasibility of scaling to large-aperture systems. Without rigorous study of these microwave, calibration, and packaging dimensions, optical beamforming networks risk remaining laboratory curiosities rather than evolving into competitive alternatives to mature electronic phased-array hardware.

### 4.3 Radar, Spectrum Sensing, and Frequency Measurement Chips

MWP radar capitalizes on several key photonic-assisted functionalities: microwave frequency multiplication, broadband microwave chirped waveform generation and de-chirping, as well as low-loss and high dynamic range microwave signal distribution. In response to the limitations of bulk optical components, which are alignment-sensitive, vibration-prone, and power-inefficient, recent research has increasingly focused on transferring these core functions from discrete, fiber-pigtailed assemblies to fully photonic integrated platforms. Two material systems have attracted particular attention: TFLN and silicon.

Building on a TFLN platform, a representative and notable advance in this direction is the work by Zhu et al., who reported an integrated lithium niobate photonic mmWave radar operating in the V band (40-50 GHz) with 1.50 cm ranging resolution, 0.067 m/s velocity resolution, and two-dimensional inverse synthetic aperture (ISAR) radar imaging with 1.50 cm × 1.06 cm resolution [14]. The architecture employs two electro-optic modulators on the same TFLN chip. The first modulator is used for broadband radar waveform generation through optical frequency multiplication and the second is used for de-chirping of the received echoes. Unlike prior demonstrations that reported standalone modulator performance, this work directly validates the electro-optic capabilities of TFLN, high linearity, broad instantaneous bandwidth, and low drive voltage, at the radar-system level, therefore it marks a step toward a fully integrated MWP radar. Fig. 4 shows the TFLN PIC which consolidates microwave waveform generation and de-chirping within a compact V-band radar front end, highlighting the potential for size, weight, and power reduction in future deployable systems.

Complementing the advances on TFLN, silicon photonics has also delivered notable progress in integrated MWP radar functionality. Ma et al. demonstrated a high-performance integrated MWP transceiver for a dual-band synthetic aperture radar (SAR) [65]. This work is particularly significant because it establishes the relationship between nonlinear mechanisms in silicon and system-on-chip radar performance through a combination of a parallel photonic-circuit architecture with waveform predistortion, linearity optimization, and thermal-crosstalk compensation. The demonstrated transceiver supports broadband C- or Ku-band radar-signal generation, stretch processing with an SFDR exceeding 100 $dB{\cdot}Hz^{2/3}$, and centimeter-level two-dimensional imaging with high peak-to-sidelobe ratios. Beyond its specific hardware achievements, the work by Ma et al. also highlights a broader principle in IMWP: chip-level radar performance depends not only on photonic circuit layout, but equally on analog impairment modeling and calibration. This systems-level perspective marks a maturation of the field, moving from device-centric demonstrations toward robust, repeatable system-on-chip engineering.

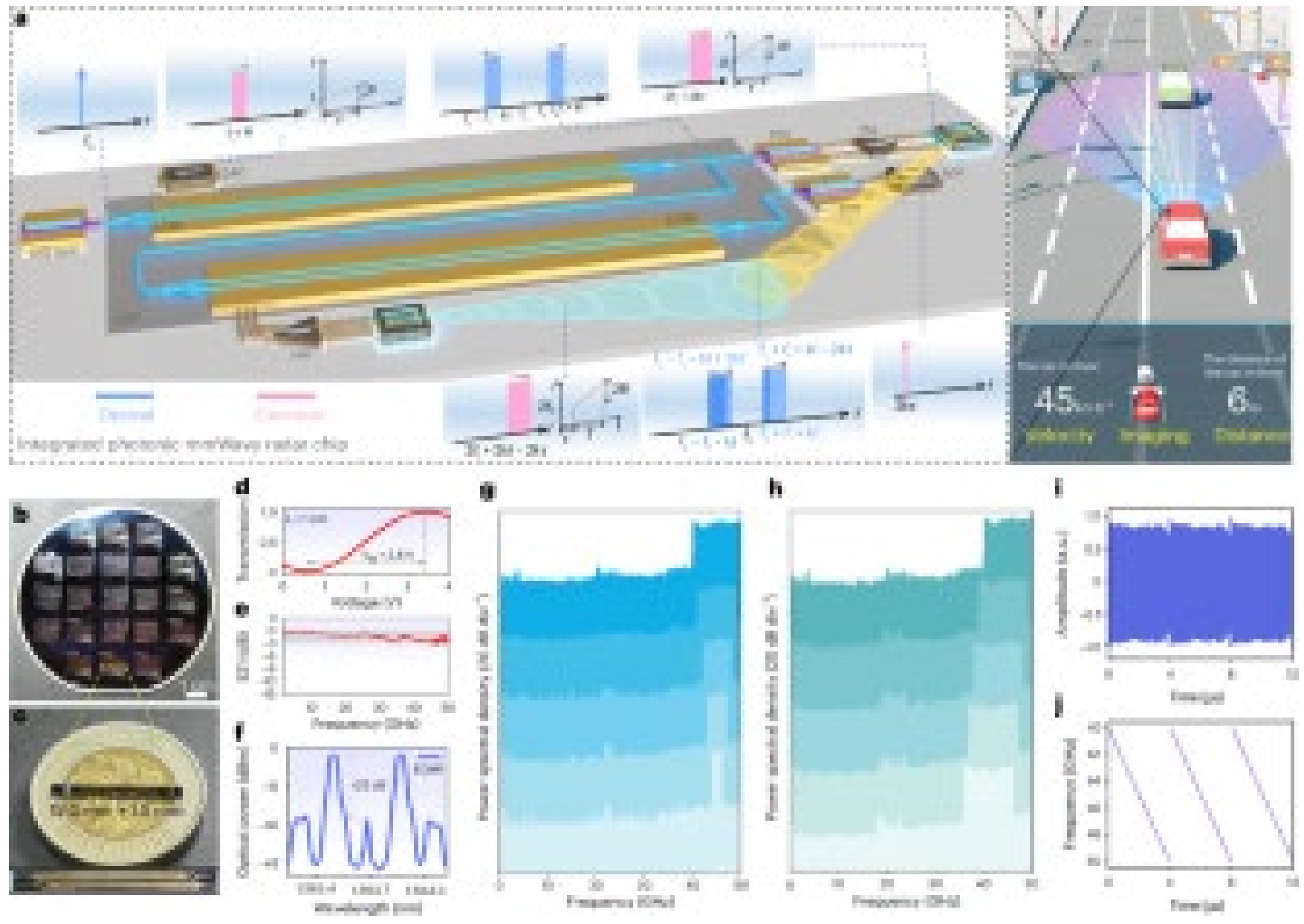

*Fig. 4. Integrated lithium niobate photonic mmWave radar chip. The figure shows a wafer-scale TFLN photonic radar layout, two electro-optic modulators for photonic-assisted microwave frequency multiplication and de-chirping, chip/wafer photographs, and measured V-band radar waveform generation. Source: Zhu et al., "Integrated lithium niobate photonic millimetre-wave radar," Nature Photonics 19, 204-211 (2025), Ref. [14], https://doi.org/10.1038/s41566-024-01608-7.*

Spectrum sensing represents another critical system function that is now transitioning from discrete optical assemblies to photonic integrated platforms. This transition is particularly timely considering the growing spectral congestion and the dynamic frequency allocation requirements foreseen for next-generation wireless networks, including 6G integrated sensing and communication (ISAC) systems. Tao et al. demonstrated a TFLN chip specifically designed for real-time ultrawideband spectrum sensing in a 6G ISAC scenario [8]. The chip integrates three key electro-optic elements on a single TFLN platform: a MZM, an electro-optic comb reference, and an electro-optic microring filter bank. This integrated architecture implements a frequency-to-time mapping function that effectively transforms the input RF spectrum into a temporal waveform, enabling rapid spectral analysis without the need for sweeping local oscillators or bulky dispersive elements. The demonstrated system achieves an analysis bandwidth of 57.5 GHz with a processing latency of less than 110 ns, a figure that is orders of magnitude faster than conventional digital spectrum analyzers or swept-tuned receivers. Fig. 5 illustrates how this chip-level sensing functionality is directly coupled with dynamic spectrum management protocols, showing the closed-loop pathway from spectrum occupancy detection to reconfiguration of communication and radar parameters, thereby highlighting the role of integrated photonics as an enabling technology for intelligent, spectrum-agile wireless systems.

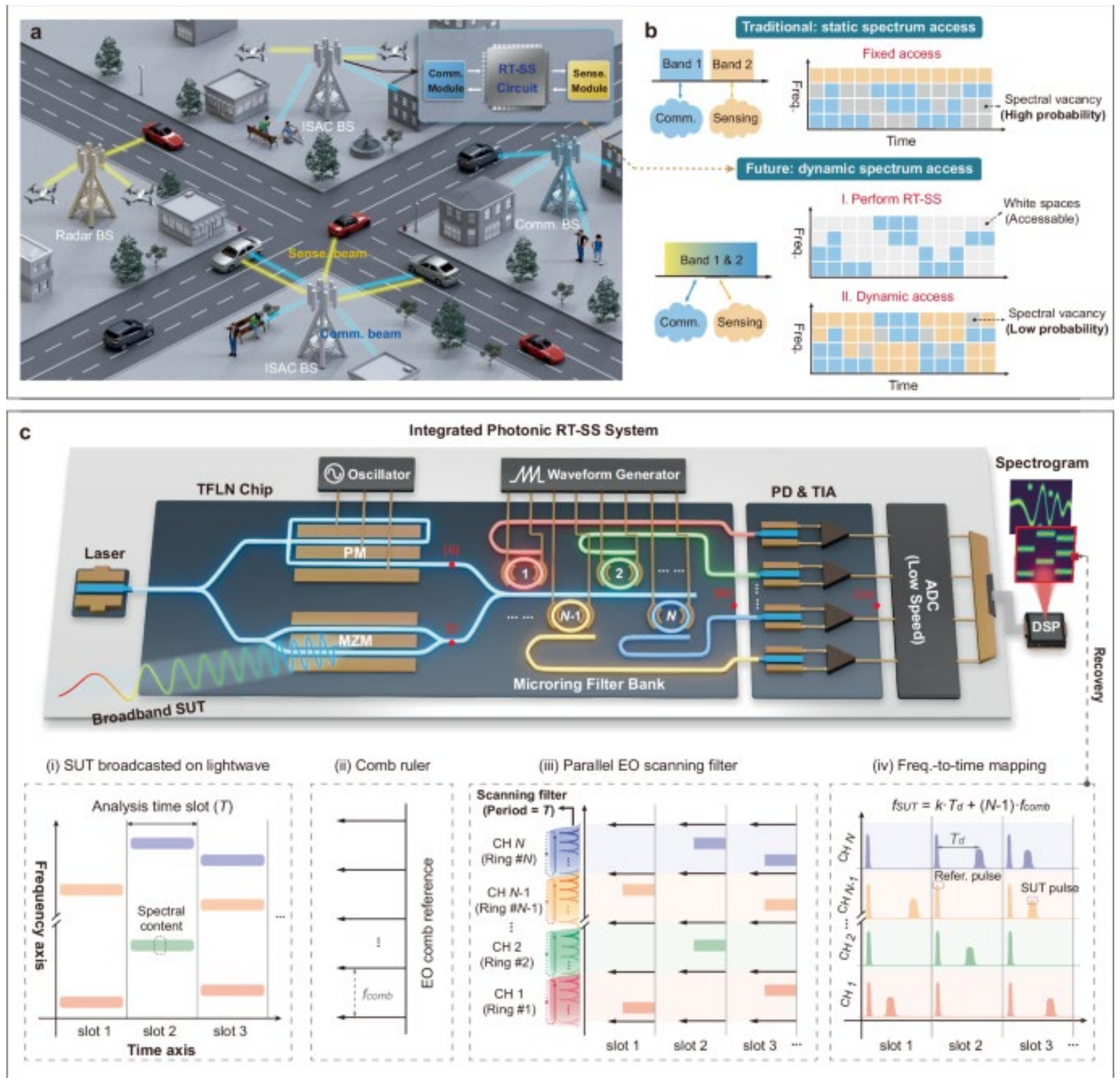

*Fig. 5. Photonic integrated real-time spectrum sensing for 6G integrated sensing and communication (ISAC). The figure shows the envisioned ISAC scenario, dynamic spectrum access concept, and the TFLN chip architecture that combines an MZM, electro-optic comb reference, and electro-optic microring filter bank for frequency-to-time mapping. Source: Tao et al., "Integrated photonic ultrawideband real-time spectrum sensing for 6G wireless networks," Nature Communications 17, 3666 (2026), Ref. [8], https://doi.org/10.1038/s41467-026-70389-0.*

Chip-scale microwave frequency measurement has rapidly emerged as a foundational primitive for modern radar and spectrum-sensing systems, fundamentally reshaping how instantaneous frequency information is acquired and processed on a PIC. This paradigm shift is driven by the pressing need for real-time spectral awareness in bandwidth-constrained and dynamically congested environments, where traditional electronic approaches struggle with speed, resolution, and size-weight-and-power (SWaP) constraints. A key trend in this evolving landscape is the progressive shift toward photonic integrated discriminators for spectral analysis, which eliminates the need for bulky dispersive elements or complex digital back-end processing. This movement toward full integration is exemplified by recent demonstrations of fully on-chip instantaneous frequency measurement architectures and highly integrated dual-modality microwave frequency identification schemes, both of which showcase the maturity and versatility of photonic integration for real-time microwave frequency measurement [66], [67]. These advances build on earlier work in photonic-assisted instantaneous frequency measurement, broadband microwave characterization, and frequency-to-power mapping, techniques that together established the core measurement toolbox for the current photonic integrated chips [68], [69]. More recently, metal patterning was used to selectively suppress unwanted high-order modes in a high-Q microdisk resonator, yielding a clean spectral response [70]. Building on this device, a photonic microwave frequency-measurement scheme aligned the ±3rd-order optical-comb sidebands with the steep slopes of two adjacent resonator notches, achieving a resolution of 4.37 kHz [71]. For this class of chips, the most instructive assessment criteria are therefore instrument-level metrics: ambiguity range, instantaneous bandwidth, calibration drift, optical power budget, and compatibility with real-world microwave receivers. Extending this measurement paradigm, a miniaturized microcomb-assisted spectrometer that leverages GHz-scale laser tuning and an integrated SiN soliton microcomb to achieve multi-THz optical analysis bandwidth while retaining high frequency precision has been reported [72].

### 4.4 Wireless and Fiber-Wireless Integrated Front Ends

Future wireless networks are expected to use sub-6 GHz, microwave, mmWave, and sub-THz bands adaptively. This coexistence creates a significant hardware challenge: deploying separate electronic front ends for each band are bulky and inefficient, yet wideband electronic frequency conversion becomes increasingly difficult at high carrier frequencies. Integrated electro-optic platforms, especially TFLN, offer a promising pathway toward broadband and reconfigurable MWP front ends.

A notable advance in monolithic integration is the work by Moller de Freitas et al., who demonstrated an ultrawideband photonic receiver fully integrated on a TFLN-on-quartz platform [31]. The principal significance of this result lies in the integration boundary itself: the receiver co-integrates antenna structures and photonic modulators on the same chip, eliminating the need for external probes or discrete interconnections between the microwave capture and optical conversion stages. Rather than connecting separately packaged components, the approach treats the entire front end from free-space microwave reception to electro-optic conversion as a single, cohesive chip. This level of integration is critical for practical ultrawideband systems, where interface parasitics and packaging constraints often degrade the system performance.

A compelling system-level demonstration of ultrabroadband on-chip photonics is provided by Tao et al., in which a TFLN photonic wireless system covering the full spectrum from 0.5 to 115 GHz was demonstrated [9]. The system integrates essential functional elements for baseband modulation, broadband wireless-photonic conversion, and reconfigurable carrier/local oscillator generation, and achieves adaptive wireless communication over nine consecutive bands with up to 100 Gbps per band. This work shows that photonics can be leveraged directly as a broadband front-end technology, capable of delivering the wide spectral coverage demanded by next-generation wireless systems, exemplified by the full-spectrum wireless system shown in Fig. 6.

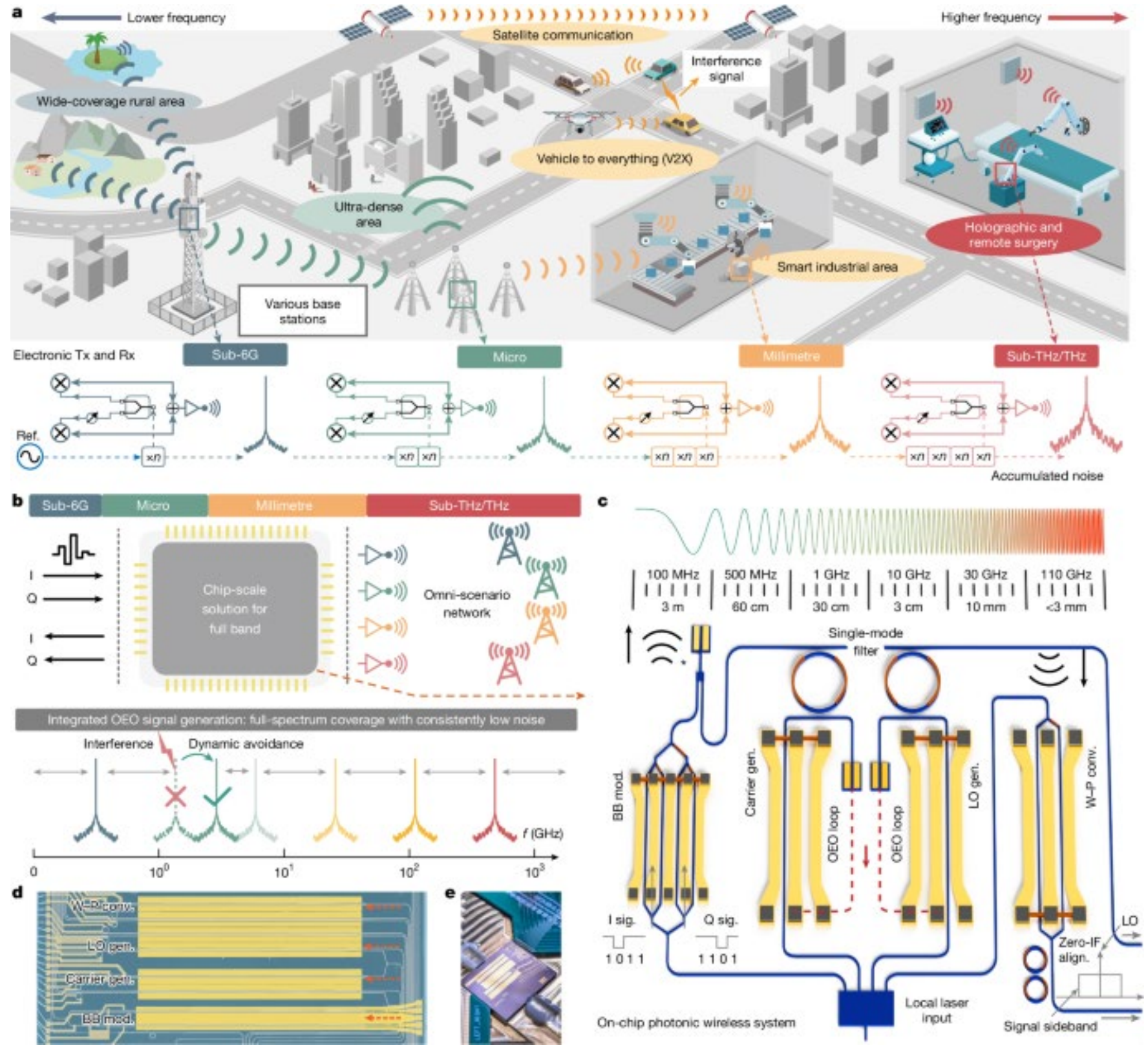

*Fig. 6. Ultrabroadband TFLN photonic wireless system for full-spectrum operation. The figure summarizes a chip-scale TFLN solution for adaptive wireless operation from low microwave to sub-THz bands, including broadband carrier/local-oscillator generation, baseband modulation, wireless-photonic conversion, and co-packaged hardware. Source: Tao et al., "Ultrabroadband on-chip photonics for full-spectrum wireless communications," Nature 645, 80-87 (2025), Ref. [9], https://doi.org/10.1038/s41586-025-09451-8.*

Another landmark demonstration of integrated photonics for ultrawideband fiber-wireless communication has been reported by Zhang et al., in which TFLN electro-optic modulator and InP-based modified uni-traveling-carrier (MUTC) photodetector were hybrid integrated to realize a system with over 250 GHz of operational bandwidth [10]. This bandwidth figure is particularly noteworthy because it simultaneously spans the entire microwave, millimeter-wave (mmWave), and sub-THz regimes, positioning the platform as a unified solution for the disparate frequency bands that next-generation wireless networks are expected to occupy. The system leverages TFLN for electro-optic conversion alongside InP-based MUTC photodetectors for photodetection, both offering an operational bandwidth exceeding 250 GHz. This hybrid integration approach enabled two notable transmission results: 512 Gbps over short-reach fiber links and 400 Gbps over wireless links. Critically, this work shows that co-designed EO and OE components, when integrated on compatible platforms, can seamlessly bridge the optical fiber domain and the high-frequency wireless domain within a single hybrid system. Such demonstrations mark a transition from individual component records toward meaningful system-level interoperability across different material platforms, a capability illustrated by the photonic integrated chip in Fig. 7, which covers both optical-fiber and THz wireless links.

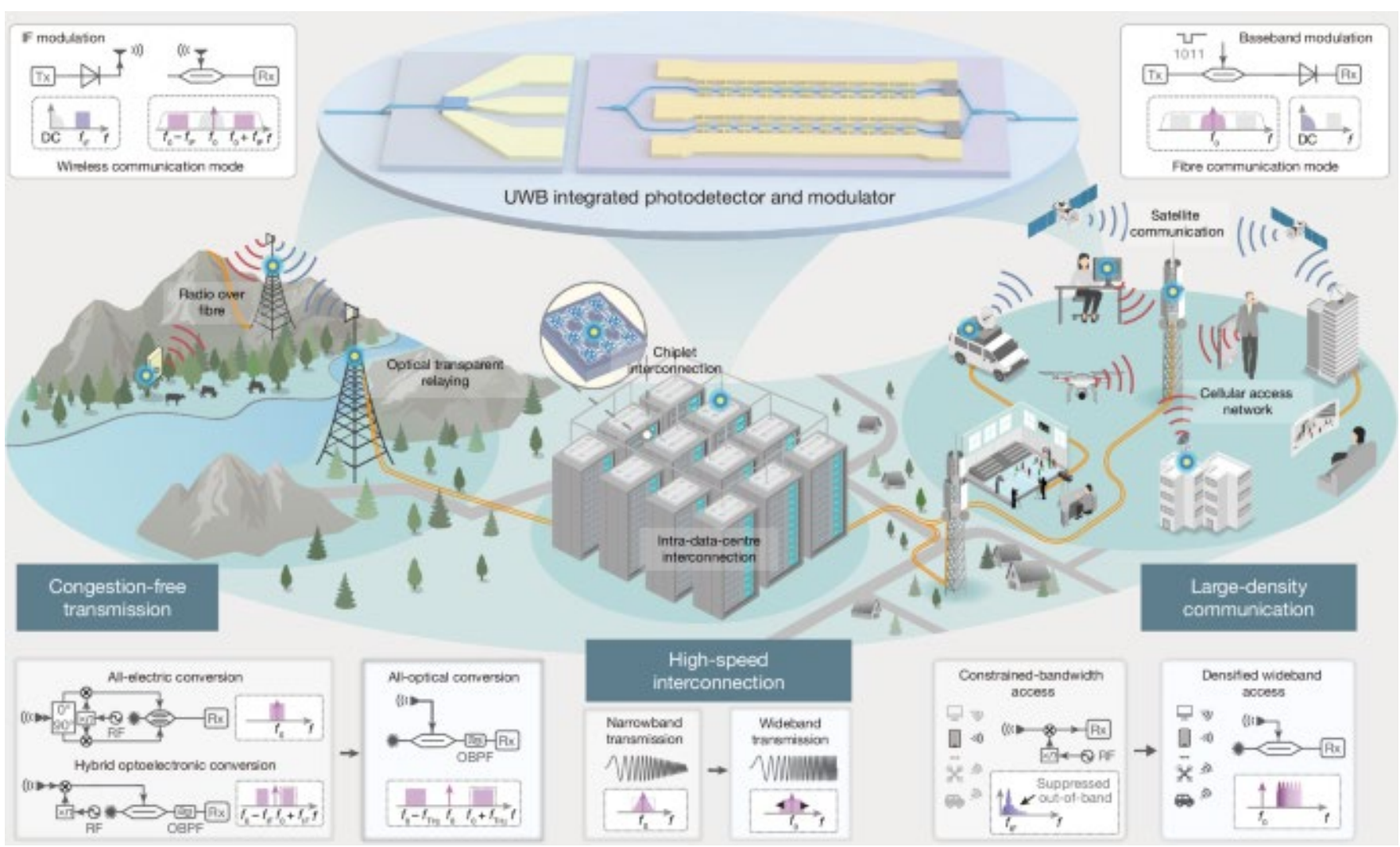

*Fig. 7. Integrated photonics for ultrawideband fiber-wireless communication. The figure illustrates an integrated ultra-wideband optical telecommunication system that uses wide-bandwidth electro-optic modulator and opto-electronic photodetector to bridge fiber interconnection and wireless access/relay scenarios. Source: Zhang et al., "Integrated photonics enabling ultra-wideband fibre-wireless communication," Nature 651, 348-355 (2026), Ref. [10], https://doi.org/10.1038/s41586-026-10172-9.*

A different and more aggressive front-end was reported by Liu et al., in which the RF aperture itself was moved into the electro-optic receiver by co-designing a bow-tie antenna with a high-Q TFLN microring modulator [73]. In this architecture, the antenna is not treated as an external microwave accessory feeding a packaged modulator; rather, it forms the RF resonant element that concentrates the incident field directly across the optical modulator. The microring simultaneously provides optical resonance, so the microwave-to-optical conversion benefits from dual RF/optical resonance rather than from a long travelling-wave electrode or a discrete antenna-LNA-EOM chain. The reported chip occupies about 2 mm × 1.7 mm and reaches a figure of merit of 3.88 $W^{-1/2}$, while packaged demonstrations include centimeter-level radar ranging, 3.2 Gbps wireless reception, and real-time high-definition video transmission. This result redefines where the integration boundary lies. Rather than confining integration to photonic signal processing functions, such as modulation, filtering, or wavelength conversion, this work extends the integration frontier all the way to the RF-photonic aperture itself. In other words, the antenna and the optical modulator are no longer treated as separate entities connected by external cabling or packaging; they are co-designed as a single resonant structure that directly captures and processes microwave signals in the optical domain. By bringing the aperture into the photonic chip, this approach eliminates the interface parasitics, impedance mismatches, and packaging losses that have historically limited the performance of discrete microwave-photonic links, while simultaneously reducing the size, weight, and power consumption, critical considerations for real-world deployment. As shown in Fig. 8, the main design idea is to align antenna resonance, near-field confinement, and optical readout on a single TFLN platform. Unlike broadband link demonstrations that still rely on external antennas and microwave packaging, this work points to conformal or miniature receivers for UAVs, high-speed trains, and distributed sensing nodes, applications where the antenna, modulator, optical carrier, and package cannot be optimized independently.

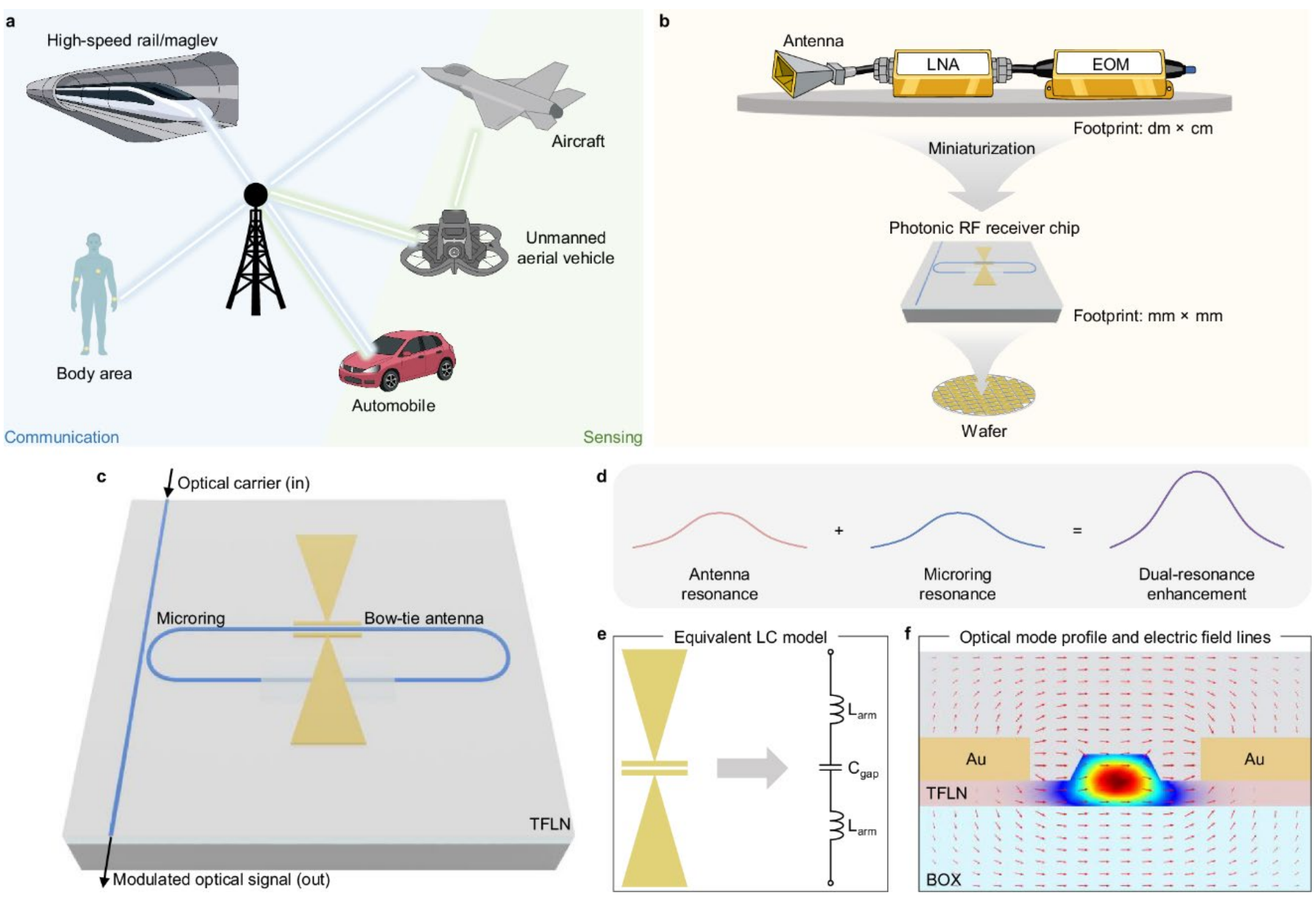

*Fig. 8. Synergistic antenna-modulator integration for a monolithic photonic RF receiver. The reproduced core figure summarizes application scenarios for photonic RF reception, miniaturization from a discrete antenna-LNA-EOM chain to a wafer-scale receiver chip, bow-tie antenna coupling to a TFLN microring modulator, dual RF/optical resonance enhancement, the equivalent LC model, and optical/electric-field overlap. Source: Liu et al., "Synergistic antenna-modulator integration for a monolithic photonic RF receiver," Nature Communications 17, 4960 (2026), Ref. [73], https://doi.org/10.1038/s41467-026-73862-y.*

### 4.5 Programmable MWP Processors and Multifunctional Processing Engines

Programmability and reconfigurability are becoming increasingly important in IMWP systems, driven by the growing demand for versatile hardware that can adapt to dynamically changing operational environments. In traditional MWP systems, each function, whether filtering, delay, beamforming, or frequency conversion, typically requires a dedicated hardware implementation, resulting in inflexible systems that cannot be repurposed without significant redesign. A programmable MWP processor, by contrast, is expected to implement multiple functions from a single common hardware platform, simply by reconfiguring its internal states through electrical or optical control. The envisioned functionality spans a wide range of signal processing operations, including filtering, delay, beamforming, frequency conversion, switching, waveform shaping, and interference suppression, all of which would be accessible on demand without hardware modification. The first demonstration of a reconfigurable multifunctional MWP signal processing chip was reported by Liu et al. using an InP/InGaAsP platform [74]. This pioneering work established the fundamental feasibility of on-chip reconfigurability, proving that the flexibility of software-defined operation could be brought to the photonic domain. It also laid the groundwork for subsequent efforts by identifying the key building blocks, tunable elements, low-loss waveguides, and fast control interfaces, necessary for reconfigurable photonic processors. Building on this foundation, later demonstrations have advanced the field along three complementary dimensions: component density has been increased to support more complex operations, the functional repertoire has been expanded to cover a wider range of signal processing tasks, and the reconfiguration speed has been accelerated to enable faster adaptation.

Perez-Lopez et al. demonstrated a general-purpose programmable photonic processor that integrates the photonic, electronic, and software stack required to configure a silicon photonic mesh for advanced microwave applications [11]. This programmable mesh represents a significant departure from fixed-function photonic circuits, as it can be reconfigured to support multiple MWP functions, including filtering, delay, and beamforming, on a common hardware substrate. The underlying concept is that the same physical chip can be dynamically reprogrammed to perform different signal processing tasks, much like a field-programmable gate array (FPGA) in the digital domain. This FPGA analogy is both instructive and aspirational, it captures the vision of a universal hardware platform that can be adapted to changing requirements without the need for physical redesign or replacement. However, the insertion loss, calibration, tuning power, and control complexity remain major challenges. Despite these challenges, the demonstration represents a critical milestone in the evolution of programmable MWP systems. Fig. 9 shows the programmable photonics approach, where the photonic core, control electronics, and software layer work together as a single integrated system.

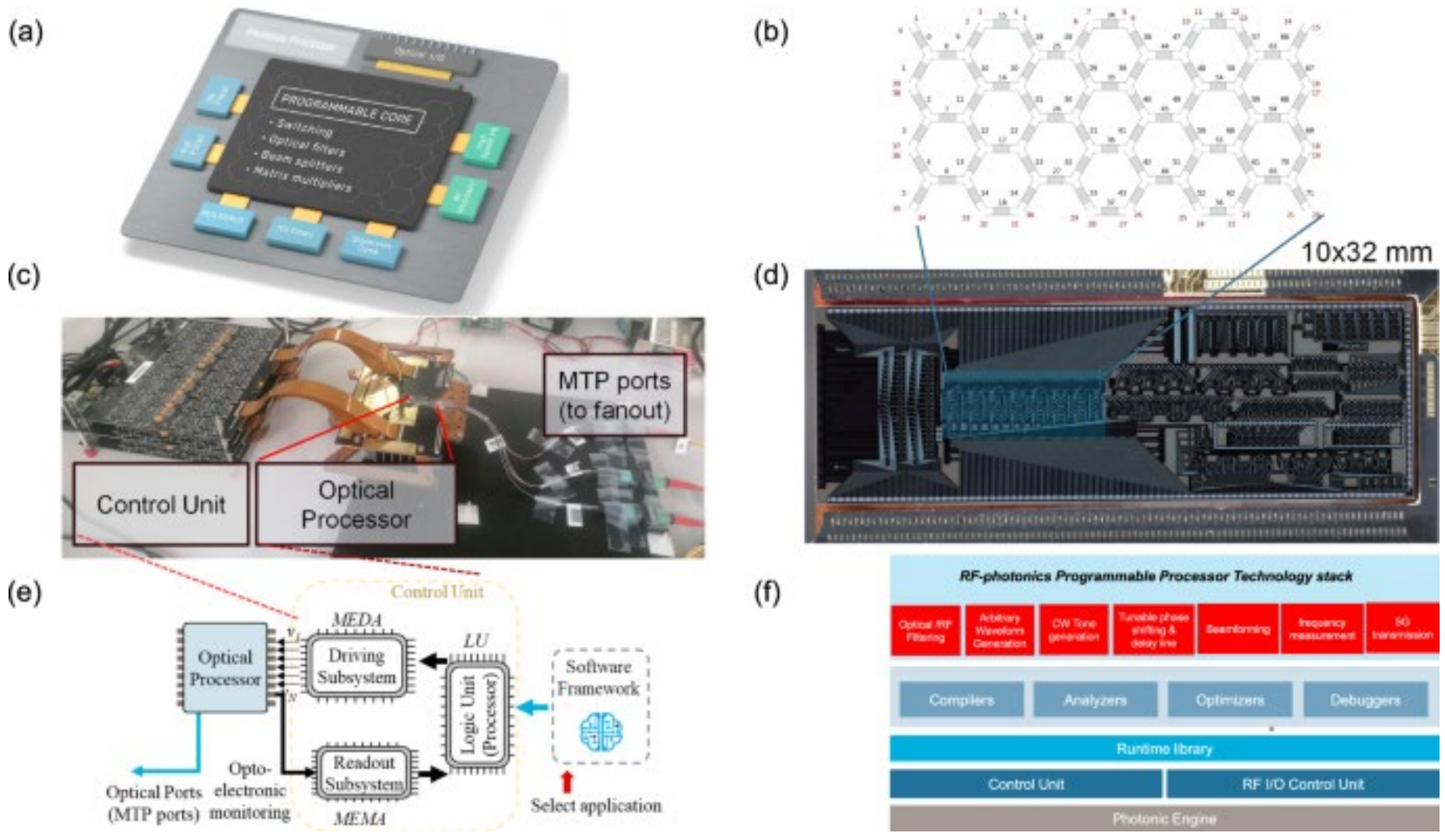

*Fig. 9. General-purpose programmable photonic processor for advanced microwave applications. The figure shows the optical layer, mesh core, packaged processor, electronic control, and software stack of a programmable silicon photonic processor. Source: Perez-Lopez et al., "General-purpose programmable photonic processor for advanced radiofrequency applications," Nature Communications 15, 1563 (2024), Ref. [11], https://doi.org/10.1038/s41467-024-45888-7.*

Photonic integrated computing provides a useful bridge between programmable optical processors and IMWP engines because many microwave signal-processing operations are linear transformations, convolutions, or adaptive weighted sums. Zheng et al. reported an SOI photonic integrated computing engine for concurrent matrix and convolution operations [16], [75], [76]. As shown in Fig. 10, the computing engine is reinforced by two related studies from the same processing engine line: an integrated optical matrix operator was used for deep residual U-Net biomedical image segmentation, while a programmable photonic integrated processor was employed for real-time microwave communication signal processing, including multipath interference compensation and self-interference cancellation over the 1.95-2.05 GHz band with up to 24.61 dB suppression. These results showed that matrix-configurable photonic processors are not limited to optical computing workloads; they can equally well support analog microwave communication functions, including multipath interference compensation and self-interference cancellation. This dual-use capability strengthens their potential for deployment in microwave front ends, where application-specific microwave interfaces, calibration, and packaging beyond the optical matrix core are needed. Adjacent work on all-optical real-time equalization for terabit-per-second links further illustrates the broader move toward low-latency photonic signal processing. More importantly, optical-link equalization and MWP signal processing share similar requirements in adaptive filtering, channel compensation, and real-time reconfigurability, suggesting that programmable IMWP chips could be extended toward more complex microwave scenarios requiring dynamic impairment mitigation and low-latency signal conditioning [77].

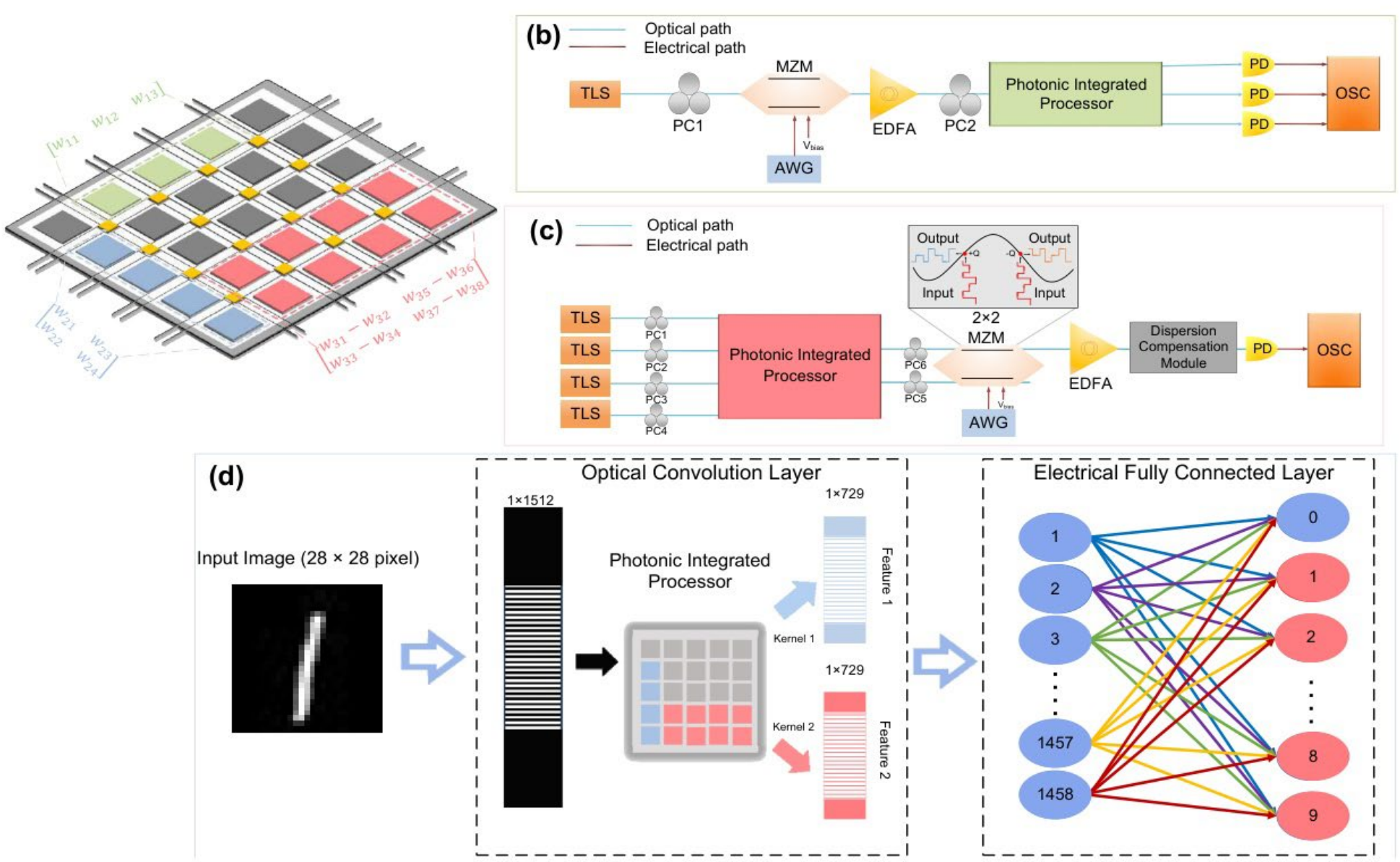

*Fig. 10. Photonic integrated computing engine for concurrent optical computing. The figure shows a segmented SOI photonic integrated processor and representative experimental configurations for parallel optical matrix and convolution tasks. Source: Zheng et al., "Photonic integrated computing engine for concurrent optical computing," Nature Communications, Aug. 2026, Ref. [16], https://doi.org/10.1038/s41467-026-76872-y.*

Recently, a single-chip silicon photonic processing chip reported by Deng et al. [12] has advanced the programmability in a different direction from the programmable mesh architecture discussed previously. Rather than focusing on reconfigurable interconnections between passive waveguide elements, this work pursues a vision of functional completeness, integrating active and passive components on a single platform to create a self-contained photonic engine. Specifically, the chip integrates lasers, high-speed modulators, a programmable optical filter, photodetectors, and switches in a single chip. This level of integration enables a diverse range of operations, including optical-to-electrical conversion, electrical-to-optical conversion, optical filtering, microwave filtering, frequency doubling, and optoelectronic oscillation all within a compact silicon photonic engine, as illustrated in Fig. 11. Unlike the programmable mesh approach, which emphasizes reconfigurable interconnections, this chip emphasizes dense co-integration of active and passive elements, including the optical source, modulator, filter, detector, optical switches, and package-level microwave access, all on a single chip.

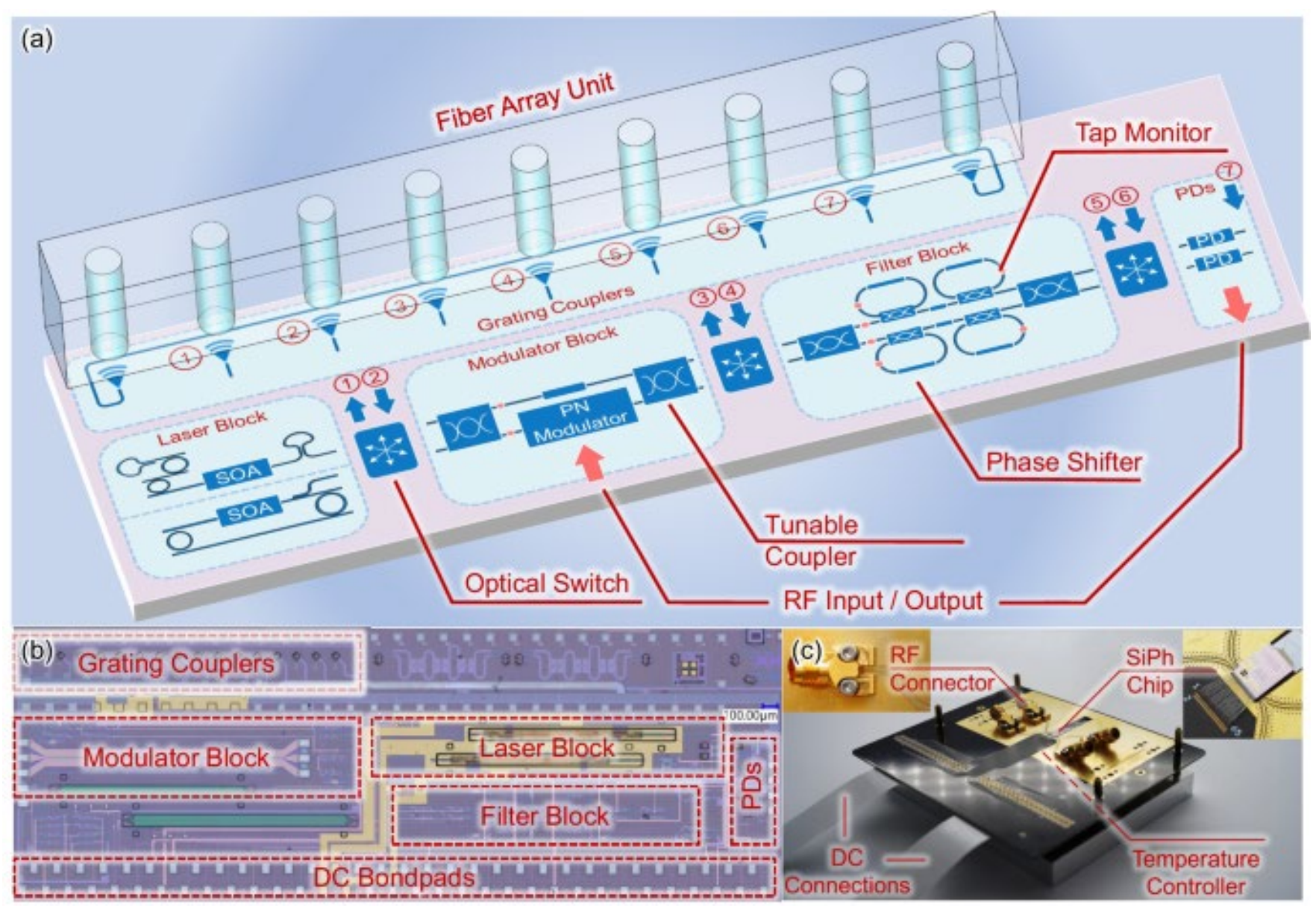

*Fig. 11. Single-chip silicon photonic engine for analog optical and microwave signal processing. The figure shows the silicon photonic integrated engine, including transfer-printed InP gain, modulator block, filter block, photodetectors, optical switches, grating couplers, and packaged demonstrator with microwave connectors. Source: Deng et al., "Single-chip silicon photonic engine for analog optical and microwave signals processing," Nature Communications 16, 5087 (2025), Ref. [12], https://doi.org/10.1038/s41467-025-60100-0.*

Programmable processing based on TFLN integrated processors is also advancing fast. Feng et al. built a TFLN MWP processing engine that integrates electro-optic modulators, and programmable optical delay and weighting units in a single chip, as shown in Fig. 12. In the system, the microwave signals are mapped to the optical domain, where signal processing was performed through parallel optical paths with engineered delays and weights. The processed signals were then converted back to the electrical domain. The integrated processor demonstrated up to 256 GSa/s operation and supports applications including differential-equation solving, microwave signal generation, and image processing [7]. Similarly, Wei et al. demonstrated a programmable multifunctional TFLN-based MWP processor that combines an intensity modulator with programmable cascaded microring resonators, achieving tunable notch filtering and interference suppression [30]. The two demonstrations reveal a coherent architectural trend: TFLN provides a high-linearity, broadband electro-optic interface for efficient microwave-to-optical conversion, while integrated resonators and delay networks serve as the core processing elements for implementing programmable microwave-photonic transfer functions. This combination enables compact, reconfigurable, and multifunctional signal-processing architectures, in which the superior electro-optic properties of TFLN are complemented by the filtering, weighting, and delay functions provided by integrated photonic structures.

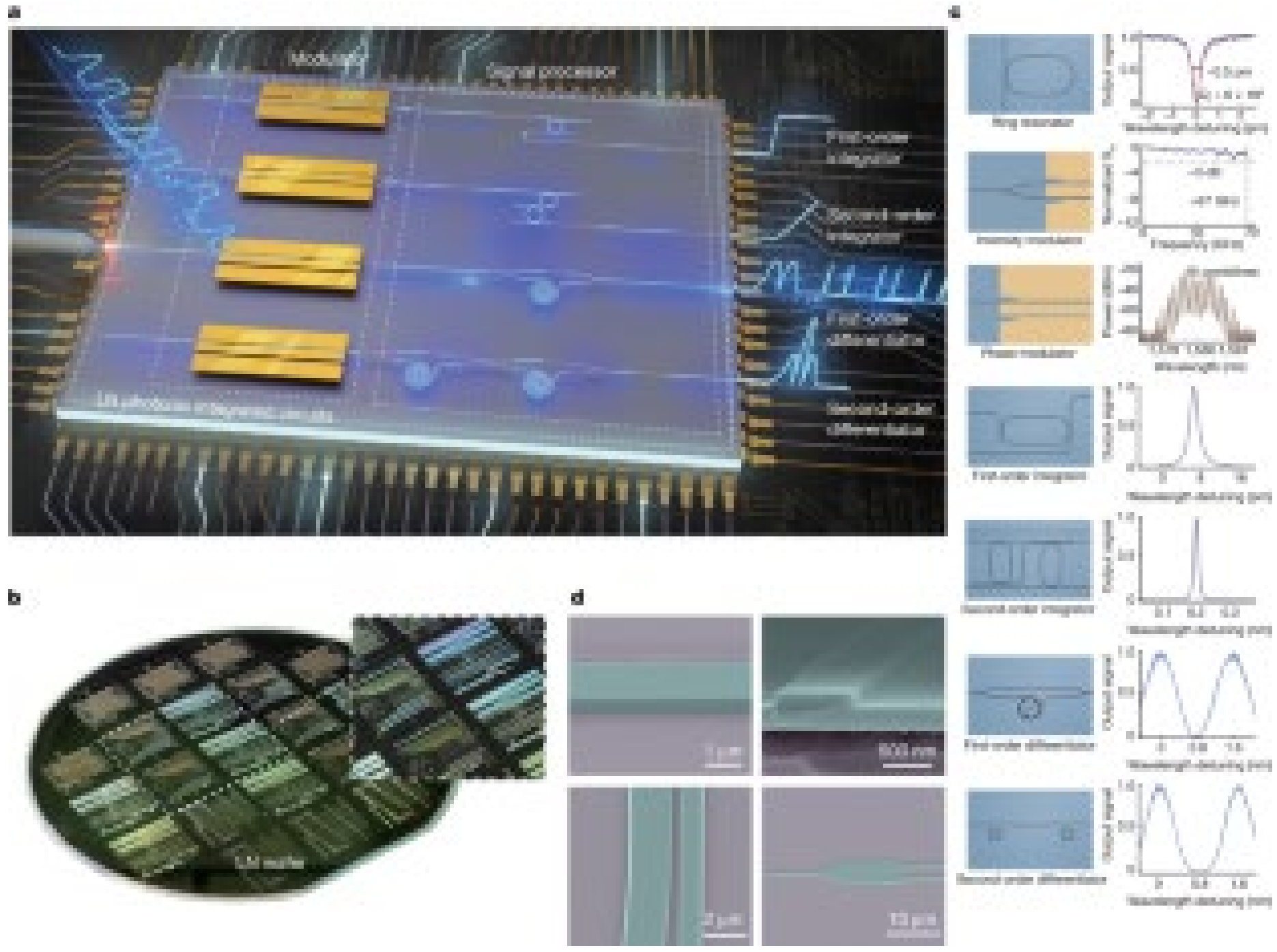

*Fig. 12. Wafer-scale TFLN MWP processing engine and building blocks. The figure shows a TFLN platform combining electro-optic modulation and functional photonic processing for ultrafast analog operations. Source: Feng et al., "Integrated lithium niobate microwave photonic processing engine," Nature 627, 80-87 (2024), Ref. [7], https://doi.org/10.1038/s41586-024-07078-9.*

Here, it is useful to distinguish universal programmability, which emphasizes broad reconfigurability across diverse functions, from function-complete specialization, which prioritizes optimized performance for a predefined class of applications. Mesh processors maximize reconfigurability, but the flexibility comes at the cost of increased insertion loss, heater count, calibration time, and control complexity. Application-specific engines can be more efficient if the target system function is known in advance. A mature IMWP ecosystem will likely need both: programmable cores that provide the flexibility needed for prototyping, adaptive processing, and multifunctional operation, and optimized engines that prioritize efficiency, performance, and compactness for established applications such as radar, spectrum sensing, beamforming, and wireless front ends. Recent demonstrations of field-programmable silicon microring WDM transceivers, non-volatile and volatile programmable optical matrices, and programmable photonic integrated meshes further demonstrate that calibration, tuning, and state retention are becoming integral system-level design considerations for programmable photonic processors, extending well beyond their traditional role as post-fabrication trimming steps [78], [79], [80].

### 4.6 Integrated Microwave Signal Generation and Compact OEOs

Microwave generation is a demanding function in MWP systems. Classical optoelectronic oscillators (OEOs) exploit a long optical delay or a high-Q optical storage element to generate low-phase-noise microwave signals [81], [82], [83], [84]. Moving toward integration, a key challenge is to determine which oscillator components can be miniaturized into a photonic chip or hybrid module without degrading the phase noise, tunability, and side-mode-suppression performance. Early photonic integrated microwave oscillators and frequency-agile photonic integrated lasers demonstrated the feasibility of replacing conventional fiber delay lines with chip-scale optical resonators, marking an important step toward compact microwave sources. At the same time, these demonstrations exposed fundamental trade-off among cavity Q factor, frequency tunability, side-mode suppression, and packaging complexity that must be carefully balanced in highly integrated oscillator architectures [85]. Tang et al. demonstrated an early InP-based integrated OEO without an off-chip fiber delay line, integrating a directly modulated laser source, an optical delay line, and a photodetector in a compact optoelectronic loop, but the short on-chip delay line led to a poor phase noise of only -91 dBc/Hz at 1 MHz offset for a 7.30 GHz carrier [86].

To improve tunability and mode selectivity, two complementary routes have been explored: tunable frequency-selective filtering [87] and parity-time- (PT-) symmetric mode selection [88], [89]. Wang et al. implemented a PT-symmetric OEO on an SOI chip by integrating an MRR-based tunable MWP filter, a PT-symmetric mode-selection circuit, and two photodetectors. The resulting OEO was tunable from 0 to 20 GHz and, at 13.67 GHz, achieved a side-mode suppression ratio of 46 dB and a phase noise of -80.96 dBc/Hz at a 10-kHz offset [90]. Zhang et al. demonstrated a compact hybrid-integrated OEO combining a laser chip, a silicon photonic chip, wire-bonded electronic chips, a compact fiber ring, and a YIG filter, in which a single-passband MWP filter with a frequency tunable range from 3 to 18 GHz was implemented, supporting microwave generation across the same frequency range with a low phase noise of -128.04 dBc/Hz at 10-kHz offset for a 10-GHz oscillation signal [91]. Li et al. reported an SOI-based integrated OEO in which an optical feedback loop and an optoelectronic feedback loop are mutually coupled through integrated modulation, filtering, coupling, and photodetection blocks, As shown in Fig. 13. The integrated OEO achieved 2-30 GHz tunable microwave generation with phase noise as low as -132 dBc/Hz at 10 kHz offset and an injection-locked Allan deviation of $10^{-12}$ at 1 s [92]. A key limitation of the reported integrated OEOs is that ultralow phase noise is often maintained by using a long off-chip fiber delay line, which compromises the compactness and robustness required in many practical applications. Future integrated OEOs should therefore focus on incorporating high-Q on-chip energy-storage elements into the oscillation loop to achieve low phase noise and compact integration simultaneously.

In addition to microwave generation, an OEO can also be employed for optical computing. For example, nonlinear OEO (NOEO)-based photonic accelerator for reinforcement learning (RL) has been reported [93]. By carefully balancing the gain and nonlinearity in the NOEO cavity, four parallel, orthogonal chaotic sequences with a 6-dB bandwidth of up to 18.18 GHz and a permutation entropy of 0.9983 were generated, providing high-speed and highly complex physical signals for computation. These sequences are then used with tug-of-war and time-differential algorithms to accelerate two RL tasks: a 512-armed multi-armed bandit problem and an intelligent Tic-Tac-Toe game. The work demonstrates that NOEO-based photonic processing can provide a fast and potentially energy-efficient hardware platform for RL, while also offering potential applications in reservoir computing and neural networks [93].

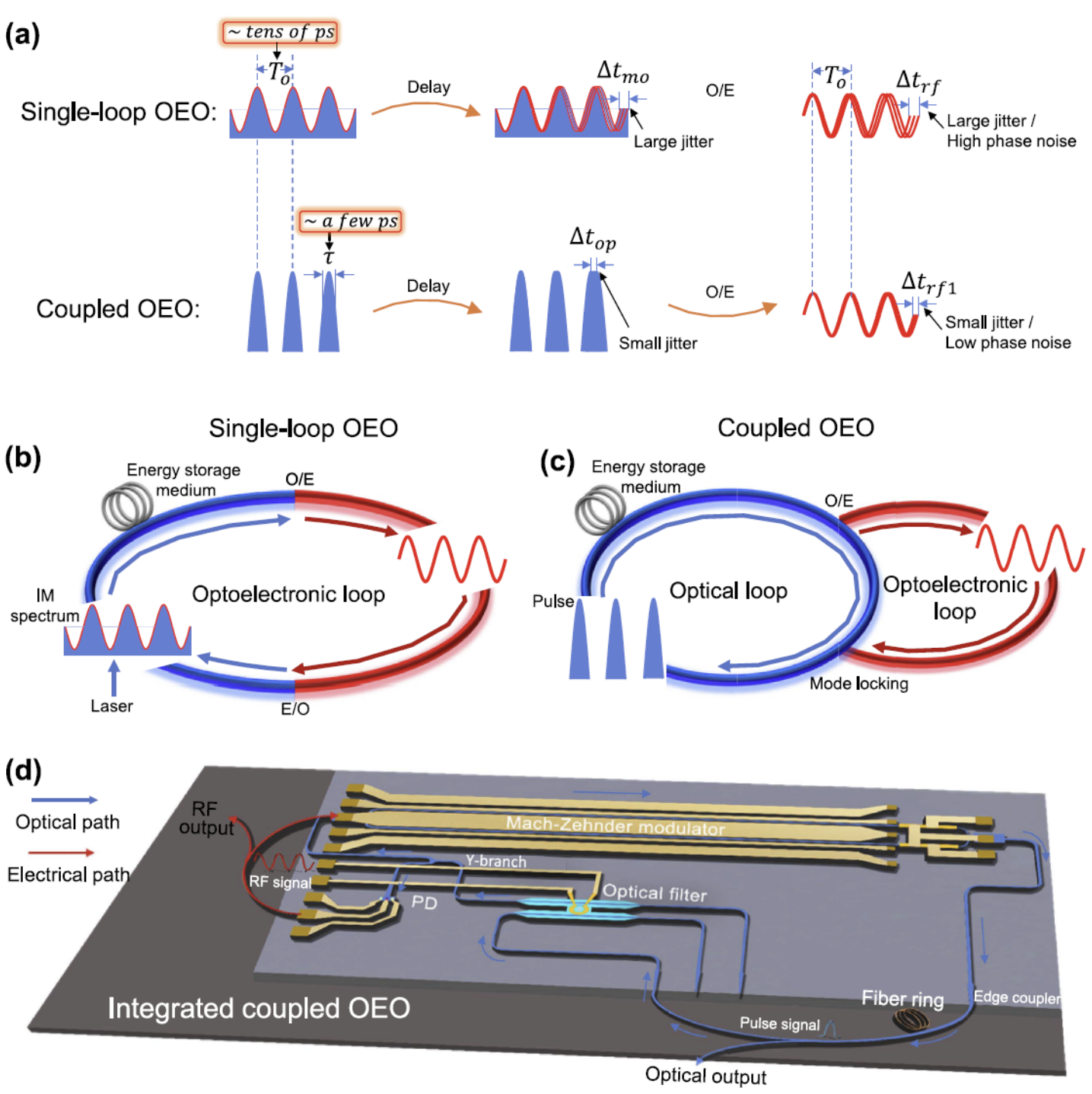

*Fig. 13. An integrated OEO using an off-chip delay fiber. The figure compares the timing jitter of the generated microwave signal between a single-loop OEO and a coupled OEO, showing the operational schematics of both configurations. Source: Li et al., "Integrated coupled optoelectronic oscillator for ultra-low-noise and ultra-wideband tunable microwave generation," Photonics Research 13, 2618-2629 (2025), Ref. [92], https:// doi.org/10.1364/PRJ.563250.*

## 4.7 Emerging Functions for Future IMWP Systems

Beyond the elementary functions already demonstrated on photonic integrated platforms, several discrete or partially photonic integrated systems are defining realistic functional targets for future IMWP hardware. These include frequency-comb-assisted true-time-delay beamforming [94], joint radar-communication-spectrum sensing [95], coherent and chip-based photonic radar [96], [97], [98], and overlapped-spectrum joint radar-communication [99], [100]. OEO-based studies further extend the design space to rapid waveform generation, stable single-mode operation, PT-symmetric dynamics, random oscillation, and dissipative microwave-photonic solitons [101], [102], [103], [104], [105], [106], [107].

Related photonic architectures add capabilities beyond conventional signal processing: optoelectronic parametric oscillators and coherent Ising machines enable optimization [108], [109], [110]; optical RF memories and ultrafast waveform synthesis provide storage and waveform reshaping [111], [112]; and multimode, tensor, diffractive, neuromorphic, and coherent photonic processors offer routes to analog computation [113], [114], [115], [116], [117], [118], [119], [120], [121], [122], [123], [124], [125]. Although these demonstrations differ in integration level, they identify functional targets for future IMWP systems and motivate the co-design of photonic cores, microwave interfaces, control circuitry, calibration routines, and packaging strategies.

## 5. Technical Challenges and Research Perspectives

The role of a given platform in IMWP is determined by more than isolated device-level metrics, as the ultimate value of a platform depends on how effectively its constituent components can be integrated to realize system-level functions. Metrics such as modulation efficiency, optical loss, detector bandwidth, or resonator Q factor provide important indications of individual device performance, but they do not, by themselves, capture the system-level advantages or limitations of a platform. Factors such as component co-integration, signal integrity, scalability, power consumption, reconfigurability, and compatibility with electronic and optical interfaces can be equally important in determining the overall system performance. Consequently, platforms are most meaningfully compared in terms of the system-level functions they can support and the performance they can deliver, rather than solely on the basis of individual component performance benchmarks.

This system-level perspective also highlights that no single material platform is likely to provide an optimal solution across the full range of functions required by practical IMWP systems. Different platforms offer complementary strengths, with some providing superior active-device performance, others enabling low-loss passive routing and filtering, and still others offering efficient electro-optic modulation or high-speed detection. Exploiting these complementary capabilities can therefore provide a more effective path toward improving overall system performance than relying on a single platform to implement all functions. The comparison indicates that many IMWP systems will be hybrid. A deployable 6G MWP front end, for example, could use an InP or heterogeneous III-V laser source for light generation and amplification, a TFLN modulator for high-linearity microwave-to-optical conversion, a SiN or SOI delay or filtering network for optical signal processing, an InP or Ge photodetector for optoelectronic conversion, and CMOS electronics for control and digital processing. Such architecture allows each platform to contribute its strongest capabilities while avoiding the compromises associated with forcing all functions onto a single material platform. Importantly, the resulting system may be delivered as a single packaged chip or module, even though multiple material platforms are integrated internally. Thus, heterogeneous or hybrid integration should not be viewed simply as an intermediate solution, but rather as a practical system-level strategy for combining complementary photonic and electronic technologies in advanced IMWP systems.

On the other hand, from an engineering perspective, a leading platform is the one that can close the application loop while maintaining acceptable yield, calibration requirements, and package complexity. SOI has the strongest pathway toward large-scale programmable PICs and foundry repeatability, high integration density, and increasingly established foundry processes. These advantages make SOI particularly attractive for complex circuits that require the integration of a large number of passive and active photonic elements with reproducible device characteristics. However, its performance in analog MWP links remains strongly influenced by the limitations of silicon-based electro-optic modulation, including modulator nonlinearity, insertion loss, and the resulting need for linearization and loss management. Consequently, while SOI offers compelling advantages in scalability, programmability, and manufacturing repeatability, achieving high-performance analog IMWP operation often requires additional circuit-level techniques and careful optimization of the optical and electrical interfaces.

TFLN offers outstanding electro-optic conversion efficiency, linearity, and bandwidth, making it particularly attractive for broadband microwave-to-optical interfaces and high-performance analog IMWP links. Its strong Pockels-effect response enables high-speed modulation with low optical loss and excellent linearity, which are especially important for demanding microwave and mmWave applications. However, TFLN has historically provided a less complete ecosystem for monolithic integration of optical sources, amplifiers, and detectors, and therefore often requires heterogeneous integration or hybrid packaging to realize a fully functional system.

InP remains one of the most capable platforms for active photonic functions, providing mature technologies for lasers, semiconductor optical amplifiers, and high-speed photodetectors. These capabilities make InP particularly attractive when optical generation, amplification, and optoelectronic conversion must be closely integrated. However, relatively higher passive propagation loss and lower circuit density compared with SOI or SiN constrain its scalability for large passive circuits and limit its suitability as a universal platform for complex photonic processing.

SiN, in contrast, is highly valuable as a low-loss passive photonic platform, particularly for optical delay lines, filtering, high-Q resonators, and frequency-comb-related functions. Its exceptionally low propagation loss and broad optical transparency make it well suited to applications requiring long optical paths, precise spectral control, or low-loss signal routing. Nevertheless, the lack of efficient native active and electro-optic devices means that SiN is rarely sufficient as a standalone platform for complete IMWP systems and is therefore particularly attractive as part of a hybrid or heterogeneous integration architecture.

For industrial systems, platform partitioning is likely to be more practical than platform replacement. Rather than seeking a single material platform capable of implementing every required function, future IMWP systems for radar, communications, sensing, and analog signal processing will likely combine complementary technologies according to their respective strengths. A representative architecture could incorporate a high-linearity electro-optic layer for microwave-to-optical conversion, an active III-V or heterogeneous layer for optical sources and detectors, a low-loss SiN or SOI layer for routing, delay, and filtering, and a CMOS layer for electronic amplification, control, calibration, and digital processing. Recent demonstrations of silicon photonic processors, TFLN-based wireless systems, and integrated fiber-wireless links indicate that such functional partitioning is already physically feasible. The more challenging question is therefore not whether these technologies can be combined, but whether the resulting multi-platform system can be manufactured at scale and subsequently calibrated, packaged, and qualified with stable microwave performance. In practical deployments, long-term stability under temperature variations, mechanical vibration, optical and electrical aging, packaging-induced drift, and other environmental stresses may ultimately determine whether a technically attractive hybrid architecture can transition from laboratory demonstration to a reliable industrial product.

The remaining challenges can be organized into five closely coupled categories: source integration, analog link performance, packaging, control and calibration software, and field validation.

(1) Optical gain and laser integration.

The absence of native optical gain remains a major bottleneck for SOI, SiN, and TFLN platforms. External lasers are adequate for laboratory demonstrations, but they compromise the SWaP advantage of IMWP and introduce additional challenges for system qualification and deployment. Heterogeneous III-V integration, micro-transfer printing, and photonic wire bonding therefore represent important pathways toward compact and scalable optical-source integration. The demonstrated single-chip silicon photonic engine shows the potential of integrating laser sources directly with the MWP signal path [12], but further work is needed to achieve low-noise, thermally stable, high-yield laser integration suitable for analog microwave systems. For practical deployment, relevant source -level metrics extend beyond output power to include linewidth, relative intensity noise, thermal tuning range, side-mode suppression, and long-term drift under packaged operating conditions.

(2) Analog link metrics

Digital optical communication metrics do not adequately capture the performance of IMWP systems. Because IMWP is ultimately intended to process or transport microwave signals, system-level evaluation should include microwave link gain, noise figure, third-order intercept point, spurious-free dynamic range (SFDR), compression behavior, phase noise, group-delay ripples, and calibration stability. Consistent reporting of these metrics under comparable operating conditions is essential for meaningful cross-platform comparisons and for distinguishing improvements in individual photonic devices from genuine advances in end-to-end microwave performance. Future studies should therefore adopt standardized microwave link metrics and measurement conditions, particularly for broadband, high-dynamic-range, and multi-tone operation. TFLN has strong advantages in electro-optic linearity, low-loss, and broadband modulation, but realizing these advantages at the system level still requires high-power photodetectors, low-noise optical carriers, efficient coupling, low loss and thermally stable packaging. In many applications, the decisive benchmark is not another photonic chip, but the electronic receiver, frequency synthesizer, or phased-array module that an IMWP system is intended to complement or replace, with established performance, cost, power consumption, and reliability.

(3) Packaging and microwave interfaces.

At mmWave and sub-THz frequencies, microwave packaging can become a dominant constraint on overall system performance. Probe-based demonstrations provide a convenient means of characterizing individual devices, but deployment requires a transition to robust packaging architectures that incorporate co-packaged RF and optical interfaces, impedance-controlled interposers, efficient thermal management, and reliable connectors or integrated antennas. The monolithic integration of antennas with TFLN photonic receivers [31] and the co-packaged system in a full-spectrum wireless system [9] indicate promising directions toward this goal. Nevertheless, manufacturable MWP packaging remains a major open challenge, particularly as operating frequencies increase and electrical parasitics, interconnect loss, thermal gradients, and mechanical tolerances become increasingly important. Packaging studies should therefore report not only insertion or transition loss, but also return loss, thermal resistance, mechanical tolerance, environmental stability, and assembly-to-assembly repeatability. Such system-level metrics are essential because packaging performance can ultimately determine whether the theoretical advantages of a photonic integrated platform are preserved when the device moves beyond a laboratory probe station into a practical system.

(4) Calibration, control, and software

Programmable IMWP systems require more than reconfigurable photonic hardware; their operations also rely on robust calibration algorithms, embedded monitoring, drift compensation, and high-speed control. Thermal tuning can provide high precision and wide tuning ranges, but it is typically slow and can introduce substantial static power consumption. Electro-optic tuning offers much faster response and therefore attractive for dynamic reconfiguration, but it may require higher drive voltages or careful co-design of the electrical and microwave interfaces. Nonvolatile tuning based on phase-change materials, MEMS, or ferroelectric domain control offers another route to reduce static power consumption and potentially minimizing the need for continuous biasing. As the number of programmable elements increases, however, the complexity of calibration and control can become as important as the underlying photonic architecture. A practical programmable IMWP processor will require an integrated software and control stack analogous to electronic design-automation tools and FPGA toolchains, encompassing circuit configuration, calibration, monitoring, optimization, fault detection, and system-level reconfiguration. The software layer should not be treated as an accessory to the photonic hardware. For large programmable meshes, filter banks, true-time-delay networks, or beamformers, calibration time, monitor placement, control-loop bandwidth, parameter drift, and recovery after power cycling are fundamental system specifications that directly determine usability, scalability, and reliability.

(5) Deployment beyond chip-scale demonstrations

The most consequential recent demonstrations have moved toward application-level systems, including TFLN radar [14], full-spectrum wireless system [9], real-time spectrum sensing [8], fiber-wireless communication [10], and silicon photonic engines [12]. The next necessary step is field-relevant validation under the environmental, reliability, and operational conditions encountered in practical deployments. For applications such as satellite communications, UAV, radar, electronic warfare, and 6G base-stations, IMWP modules and systems must be evaluated not only for their photonic and microwave performance, but also for thermal cycling, vibration, radiation, power consumption, manufacturability, packaging reliability, and long-term calibration stability. In this context, the system boundary defined here becomes practically important: a chip-scale demonstration should clearly specify which functions are to be integrated within the package, which remain external, and what assumptions are made regarding optical sources, electrical drivers, amplifiers, antennas, thermal management, and digital signal processing or correction. Such a clear definition is essential for assessing the true level of integration and for making meaningful comparisons of system-level performance, SWaP, cost, and deployment readiness.

(6) Recommended reporting metrics

To make IMWP a mature engineering field, future studies need to report enough information for system-level comparison. For modulators, bandwidth and $V\pi$ alone are not sufficient; optical insertion loss, microwave loss, chirp, bias stability, linearity, and power handling, all affecting system-level performance, should be studied and reported. For photodetectors, the 3-dB bandwidth should be accompanied by the saturation current, output microwave power, responsivity, dark current, and relevant impedance environment. For filters and delay lines, group delay ripples, passband insertion loss, tuning power, thermal drift, and calibration time should be reported. For complete systems, the relevant microwave-link metrics including link gain, noise figure, SFDR, EVM, BER, phase noise, latency, and long-term drift should be evaluated. The metric framework builds on established analog optical-link practice [45], [46], reporting conventions for filter and delay-lines [47], [48], [49], [50], [51], [52], [53], high-speed photodetector characterization [24], [25], and recent system demonstrations that report application metrics alongside photonic performance [8], [9], [10], [12], [13], [14], [65]. Consistent reporting of these metrics would make it possible to distinguish improvements in individual devices from genuine advances in end-to-end system performance. Without such information, it remains difficult to assess whether a reported chip represents a laboratory proof of concept or has a credible path toward deployable IMWP hardware.

(7) Open problems in IMWP systems

Five challenges remain especially important for the development of IMWP systems. The first is low-noise on-chip optical generation, for which heterogeneous or hybrid III-V integration and micro-transfer printing are two of the most active routes [12], [57], [58], [126]. The second is high-power photodetection beyond 100 GHz while maintaining low noise, high linearity, and a manageable thermal load [24], [25]. The third is the integration of low-loss, high-linearity modulators with dense programmable filters, a challenge that links the strong electro-optic performance of TFLN to silicon with the scalability and passive functionality of silicon and SiN platforms [7], [11], [27], [28], [29], [30]. The fourth is packaging: microwave transitions, optical coupling, thermal management, mechanical stability, and electronic control must be co-designed rather than optimized independently [9], [10], [31]. The fifth is the control and calibration software: a large IMWP engine may contain hundreds of heaters, phase shifters, microrings, and switches, without automated calibration, drift compensation, fault detection, and efficient reconfiguration, such hardware will remain difficult to operate reliably outside controlled laboratory environments [11], [64], [74], [127].

These challenges also define significant research opportunities. The long-term potential of IMWP lies not necessarily in replacing one platform with another, but in combining the complementary strengths of multiple technologies within a manufacturable system, i.e., a heterogeneous or hybrid platform that brings together TFLN-based electro-optic conversion, InP-based light generation and amplification, SiN-based low-loss passive routing and filtering, SOI-based integration density and programmability, and CMOS-based electronic control in a manufacturable package. Such an architecture would provide a practical pathway toward highly integrated, scalable, and reconfigurable IMWP systems for spectrum-agile communications, radar, sensing, and analog computing.

## 6. Discussion and Conclusion

IMWP is entering a system-driven phase. Early studies in the field relied on discrete photonic components to demonstrate the feasibility and value of photonics for implementing individual functions such as microwave generation, processing, distribution, and measurement. These demonstrations established the fundamental advantages of photonic approaches, including low-phase-noise microwave signal generation, broadband and low-loss signal distribution, and wideband and tunable microwave signal processing in the optical domain. The field subsequently progressed toward application-specific systems, such as true-time-delay beamforming networks and MWP radar systems, which were initially implemented using discrete components and have more recently evolved toward PICs. Particularly, over the past five years, this evolution has accelerated toward more complete chip- and module-level MWP systems. TFLN has emerged as a central platform for ultrabroadband and highly linear electro-optic conversion; silicon remains essential for large-scale integration and programmable routing; InP provides critical capabilities for light generation, optical gain, and high-speed photodetection; and heterogeneous or hybrid integration is becoming a practical approach for combining these complementary functions.

As integration progresses, the performance of IMWP systems can no longer be judged primarily by isolated device metrics. A low-$V\pi$ modulator, low-loss delay line, high-bandwidth photodetector, or high-Q resonator is valuable only insofar as it contributes to the performance of a complete microwave function. For application-oriented systems, the relevant question is whether the photonic integrated technologies can collectively provide a measurable advantage in radar, communications, sensing, frequency measurement, signal processing, or analog computing. A wider bandwidth or lower propagation loss, for example, may provide little system-level benefit if it is accompanied by excessive optical loss, poor linearity, high tuning power, difficult calibration, or packaging complexity. Conversely, a platform with less exceptional individual device metrics may offer greater overall value through dense integration, programmability, reliable manufacturing, efficient control, and robust operation.

This system-level perspective calls for more consistent reporting of complete system metrics. Future demonstrations should, where relevant, report link gain, noise figure, linearity, SFDR, phase noise, group-delay characteristics, calibration time, power consumption, and long-term drift, together with a clear definition of the functions integrated within the chip or package and the assumptions made about external components. Such reporting would enable more meaningful comparisons across material platforms and help distinguish genuine system-level advances from improvements confined to individual devices. It would also provide a more realistic basis for comparison with incumbent electronic solutions, which remain the relevant benchmark for many practical applications.

Recent demonstrations in space-ground connectivity, full-spectrum wireless systems, integrated radar and sensing, fiber-wireless communications, and concurrent optical computing further illustrate this transition [7], [8], [10], [13], [14], [15], [16]. They point toward application-oriented photonic engines in which generation, modulation, routing, filtering, detection, and electronic control are increasingly coordinated within a common system architecture. Such systems are unlikely to rely on a single material platform. Instead, TFLN can provide highly linear and broadband electro-optic conversion, InP can supply optical generation, gain, and high-speed optoelectronic conversion, SiN can provide low-loss passive routing, delay, and filtering, SOI can enable dense integration and high programmability, and CMOS can provide electronic control, calibration, and signal processing. Heterogeneous or hybrid integration and advanced packaging will therefore be important not simply for increasing the level of integration, but for combining these complementary capabilities while maintaining acceptable SWaP, manufacturing yield, reliability, and cost.

The remaining challenge is to translate these increasingly integrated architectures from well-controlled laboratory demonstrations into reproducible and deployable systems. For applications such as 6G communications, radar, electronic warfare, satellite systems, UAVs, sensing, and analog computing, IMWP systems must maintain their microwave and photonic performance under realistic thermal, mechanical, electrical, and environmental conditions. Packaging, optical-source stability, microwave interfaces, calibration and control, thermal management, and long-term drift must consequently be considered as integral parts of the system rather than as secondary implementation details. The ultimate benchmark is therefore not another isolated record in device performance, but a manufacturable and application-ready IMWP system whose microwave performance, stability, SWaP, calibration burden, reliability, and cost provide a compelling advantage over established electronic architectures. Achieving this goal would mark the transition of IMWP from an integration technology for photonic devices to a practical hardware platform for next-generation microwave systems.

**Availability of data and materials**: Data sharing is not applicable to this review article as no datasets were generated.

**Competing interests:** The authors declare no competing interests.

**Funding**: This work was supported by the National Key Research and Development Program of China (Grant No. 2024YFB2807400).

**Authors' contributions:** J.Z. and J.Y. jointly conceived the scope and overall structure of the paper. J.Z. conducted the literature survey and wrote the original draft. J.Y. critically reviewed and revised the manuscript. Both authors read and approved the final version.

**Authors' information:**

**Jiejun Zhang** received his Ph.D. degree in Electrical Engineering and Computer Science from the University of Ottawa, Canada, in 2017. He subsequently continued his postdoctoral research in the Microwave Photonics Research Laboratory at the University of Ottawa. From 2017 to 2018, he was a photonic engineer at Ciena Corporation in Ottawa, Canada. In 2018, he joined Jinan University, Guangzhou, China, where he is a professor and Director of the Jinan Photonic Integration Center. Jiejun Zhang has authored or coauthored more than 100 scientific publications, including over 60 peer-reviewed journal articles published in journals such as Science Advances, Nature Communications, and Light: Science & Applications. He also holds 18 patents related to microwave photonics and integrated photonics. He is an Associate Editor of IEEE Microwave and Wireless Technology Letters and serves on the technical committees of several international conferences, including the Asia Communications and Photonics Conference (ACP), the Conference on Lasers and Electro-Optics Pacific Rim (CLEO-PR), the Photonics and Electromagnetics Research Symposium (PIERS), and Microwave Photonics Technology and Applications (MPTA). He is a Senior Member of IEEE.

**Jianping Yao** is a Distinguished University Professor in the School of Electrical Engineering and Computer Science, University of Ottawa, Canada. Prior to joining the University of Ottawa in 2001, he was an Assistant Professor in the School of Electrical and Electronic Engineering, Nanyang Technological University, Singapore. Jianping Yao has established himself as a seminal contributor to Microwave Photonics, authoring over 400 refereed journal papers and 300 conference papers, with more than 34,000 citations and an h index of 94. He was Editor-in-Chief of IEEE Photonics Technology Letters (2017–2021), an elected member of the Board of Governors of the IEEE Photonics Society (2018–2021) and is currently a member of the IEEE Photonic Society Publication Council. He was an IEEE Distinguished Microwave Lecturer (2013–2015). Jianping Yao is a recipient of several prestigious awards, including the 2018 IEEE R.A. Fessenden Silver Medal and the 2025 IEEE Microwave Application Award. He is a Fellow of IEEE, Optica (formerly Optical Society of America), the Chinese Society for Optical Engineering, the Canadian Academy of Engineering, and the Academy of Science of the Royal Society of Canada.